\PassOptionsToPackage{table,xcdraw}{xcolor}
\PassOptionsToPackage{table,xcdraw}{xcolor}
\documentclass[final,3p,times]{elsarticle}
\usepackage[normalem]{ulem} 

\usepackage{booktabs}
\usepackage{tabularx}
\usepackage{multirow}
\usepackage{pifont}
\usepackage{array}
\usepackage{makecell}

\newcommand{\cmark}{\ding{51}} 
\newcommand{\xmark}{\ding{55}} 
\newcommand{\pmark}{--}        

\newcolumntype{Y}{>{\centering\arraybackslash}p{0.85cm}} 
\newcolumntype{Z}{>{\raggedright\arraybackslash}X}       

\usepackage{hyperref}
\hypersetup{colorlinks=true, linkcolor=blue, urlcolor=cyan, citecolor=blue}
\AtBeginDocument{\hypersetup{pdftitle={Cost-benefit analysis of an AI-driven operational digital platform for integrated electric mobility, renewable energy, and grid management},pdfauthor={Arega Getaneh Abate, Xiao-Bing Zhang, Xiufeng Liu, Dogan Keles}}}
\usepackage{booktabs}
\usepackage{amssymb}
\usepackage{amsmath}
\usepackage{amsfonts}
\usepackage{graphicx}
\usepackage{multirow}
\usepackage{booktabs}
\usepackage{longtable}
\usepackage{float}
\usepackage{array}
\usepackage{threeparttable}
\usepackage{caption}
\newcolumntype{P}[1]{>{\raggedright\arraybackslash}p{#1}}
\usepackage{algorithm}
\usepackage{algpseudocode}

\usepackage{booktabs}
\usepackage{multirow}
\usepackage{siunitx}
\usepackage[utf8]{inputenc}
\usepackage{pifont}
\usepackage{subcaption}

\usepackage{makecell}
\usepackage{empheq}
\usepackage{natbib}
\usepackage{appendix}
\usepackage{eurosym}
\usepackage{url}
\usepackage{float}
\usepackage[utf8]{inputenc}
\usepackage{textcomp}
\usepackage[T1]{fontenc}
\biboptions{numbers}
\usepackage{lmodern}
\usepackage{pdflscape}

\usepackage{tikz}
\definecolor{lightgreen}{RGB}{204,255,204}
\usetikzlibrary{matrix}
\usetikzlibrary{arrows.meta, positioning, fit, backgrounds}
\usepackage{setspace}
\usepackage{colortbl}
\definecolor{perception}{RGB}{255,242,204}
\definecolor{data}{RGB}{255,204,204}
\definecolor{middleware}{RGB}{204,255,255}
\definecolor{business}{RGB}{255,204,255}
\definecolor{application}{RGB}{204,255,204}
\usepackage{tikz}
\usetikzlibrary{positioning, fit, arrows.meta, calc, shapes.geometric,
                 backgrounds, decorations.pathreplacing}
\usepackage{helvet}

\definecolor{percC}{HTML}{1B5E20}     
\definecolor{percL}{HTML}{E8F5E9}
\definecolor{dataC}{HTML}{0D47A1}     
\definecolor{dataL}{HTML}{E3F2FD}
\definecolor{aiC}{HTML}{BF360C}       
\definecolor{aiL}{HTML}{FBE9E7}
\definecolor{appC}{HTML}{4A148C}      
\definecolor{appL}{HTML}{F3E5F5}
\definecolor{extC}{HTML}{E65100}      
\definecolor{extL}{HTML}{FFF3E0}
\definecolor{bcC}{HTML}{00695C}       
\definecolor{bcL}{HTML}{E0F2F1}
\definecolor{txt}{HTML}{263238}
\definecolor{grayA}{HTML}{90A4AE}

\journal{Applied Energy}

\begin{document}

\begin{frontmatter}

\title{Cost-benefit analysis of an AI-driven operational digital platform for integrated electric mobility, renewable energy, and grid management}

\author{Arega Getaneh Abate\corref{cor1}}
\ead{ageab@dtu.dk}

\author{Xiao-Bing Zhang}
\ead{xiazhan@dtu.dk}

\author{Xiufeng Liu}
\ead{xiuli@dtu.dk}
\author{Dogan Keles}
\ead{dogke@dtu.dk}

\cortext[cor1]{Corresponding author}

\address{Department of Technology, Management and Economics, Technical University of Denmark, Kgs. Lyngby, Denmark}

\begin{abstract}
Integrating electric mobility, including electric vehicles (EVs), electric trucks (ETs), and renewable energy sources (RES) with the power grid is paramount for decarbonization, efficiency, and stability.
A critical gap remains, however: existing smart-grid and e-mobility cost-benefit analysis (CBA) approaches do not yet provide a unified framework for appraising AI-driven operational digital platforms (ODPs) that jointly coordinate EV/ET charging, renewable generation, and grid operations across sectoral and national boundaries.
This paper develops a seven-step CBA framework tailored to this class of platform.
The framework maps each layer of a multi-layered AI architecture to traceable, monetizable benefit streams --- spanning economic efficiency, grid reliability, and environmental externalities --- while explicitly accounting for AI-specific capital and operational expenditures that conventional appraisals omit. Applied to a ten-year, three-country deployment across Austria, Hungary, and Slovenia, the analysis indicates a robust positive investment case under the modeled assumptions, confirmed through scenario sensitivity analysis, one-way parameter ranking, and probabilistic simulation. Benefit composition and country-level drivers differ systematically across national contexts, yet the economic rationale is preserved in each, reflecting the framework's adaptability to heterogeneous electrification trajectories. The findings indicate the economic viability of AI-driven digital platforms for cross-sectoral energy--mobility integration and highlight the critical role of ODPs in advancing decarbonization in the mobility--power nexus. To that end, they have direct implications for the design and appraisal of digital infrastructure investments under the EU's Fit for 55 and REPowerEU programmes.
 
\end{abstract}

\begin{keyword}
Operational digital platform \sep Cost-benefit analysis \sep AI-driven optimization \sep Renewable energy integration \sep Transport electrification
\end{keyword}
\end{frontmatter}

\section{Introduction}\label{sec_intro}
The European Union is accelerating the shift to sustainable energy and transport. Electric vehicles (EVs) and electric trucks (ETs) are achieving rapid market penetration. At the same time, the share of renewable energy sources (RES) in the grid is rising sharply \citep{motlagh2025review}. Coordinated infrastructure requirements are also expanding materially \citep{borjesson2025costs}. Ref. \cite{motlagh2025review} documents the operational complexities that accompany this transition, while \cite{borjesson2025costs} quantifies the scale of coordinated infrastructure investment it requires. The EU targets EVs/ETs to account for 60\% of new vehicle sales by 2030 \citep{mkadziel2023energy}.
Integrating EV/ET charging demands with intermittent RES and legacy grid infrastructure presents substantial technical and economic challenges \citep{psarros2023generation}. Coordinating multiple asset classes under a unified dispatch logic adds a further layer of complexity \citep{yuvaraj2024comprehensive}. Ref. \cite{psarros2023generation} demonstrates that unmanaged charging peaks impose severe stress on distribution networks. \cite{yuvaraj2024comprehensive} further traces the difficulties of coordinating multiple asset classes under a unified dispatch logic. Key challenges include the spatiotemporal mismatch between variable renewable supply and dynamic charging demand, the need for enhanced grid flexibility to accommodate high-impact loads, and the absence of coordinated platforms spanning both the energy and transport sectors.

Transportation electrification is a direct route from renewable energy integration to a zero-carbon economy. However, realizing this route requires resolving several interdependent technical bottlenecks. These include accurate battery parameter estimation in V2G systems \citep{khalid2021parameter}, grid-integrated coordination of renewables and battery storage \citep{R2_ref2_gridintegrated}, and scalable strategies for hydrogen and battery storage integration \citep{gulraiz2025energy}. Traditional approaches handle these challenges in silos, leading to suboptimal asset utilization, grid congestion, renewable energy curtailment, and underutilization of emerging flexibility resources.

The coordination challenge is particularly acute for heavy-duty transport. Battery-electric propulsion for long-haul freight is technically viable but economically constrained by battery weight, range limitations, and sparse charging infrastructure \citep{ccabukoglu2018battery}. Related review evidence points to similar practical deployment constraints \citep{cunanan2021review}. Ref. \cite{cunanan2021review} shows that optimized logistic scheduling is essential to overcome these range constraints in practice. Comparisons of battery trucks against electric road systems (ERS) further reveal the complexity of cross-border infrastructure investment decisions \citep{borjesson2025costs}. Related breakeven analysis reaches a similar conclusion \citep{deshpande2023breakeven}. To this end, Ref. \cite{borjesson2025costs} demonstrates that, without supranational coordination, individual countries may rationally forgo ERS investments even when doing so reduces aggregate EU welfare. This underscores the need for integrated digital solutions capable of bridging national and sectoral divides.

The challenges outlined above raise a concrete investment question: can an AI-driven operational digital platform (ODP) integrating mobility, energy, and grid management deliver sufficient economic, reliability, and environmental value to justify its capital and operational cost?

Answering this question requires a \textit{holistic ODP}—defined here as a functional, interoperable platform that simultaneously manages (i) EV/ET charging under fleet duty constraints, (ii) RES forecasting and curtailment reduction, (iii) grid congestion mitigation and reliability services, and (iv) cross-border market participation. Such a platform is not an offline evaluation tool. It is active operational infrastructure, maintaining time-series databases for real-time telemetry, relational databases for asset registries, and live data pipelines connected to weather APIs, energy market feeds, and grid constraint data. Existing CBA methodologies do not yet provide a platform-specific procedure for appraising such systems with explicit AI cost categories and cross-sector benefit attribution. This gap is what the present paper addresses.

Cost-benefit analysis (CBA) provides a systematic framework for evaluating the economic, financial, and societal viability of complex integrated solutions like the one proposed herein. In European projects, standardized CBA methodologies have been developed to assess smart grid and sustainable energy projects. Notably, the European Commission Joint Research Centre (EC JRC) \citep{ECJRC2012} outlines a multi-step CBA approach tailored for smart grid investments. Conversely, \cite{Stromsather2018} details a similar framework applied to a smart grid pilot project in Isernia, Italy, focusing on distribution system automation. Both methodologies advocate a sequential analytical process, encompassing scope definition, benefit identification, monetization, and uncertainty assessment.

Applying CBA frameworks to solutions that simultaneously manage energy and transport systems requires careful treatment of unique benefit streams and cost structures that standard smart-grid methodologies do not address. Existing frameworks provide a solid foundation for single-infrastructure projects, yet none rigorously quantifies the specific contributions of AI functionalities or the compounded synergies of cross-sectoral and cross-border integration in a unified platform context. Recent evaluations of integrated e-mobility systems remain largely confined to national boundaries \citep{de2023electric} or siloed sectors \citep{greaker2022economic}. Studies on V2G economic viability \citep{greaker2022economic} or real-time EV optimization \citep{feng2013economic} capture partial benefits. More recent scheduling work reaches a similar conclusion \citep{VazquezCanteli2019}, but still omits the compounded value that emerges when mobility, RES, and grid operations are optimized jointly. Existing CBA frameworks do not yet explicitly link modern AI modules to measurable KPI deltas and monetize those deltas across cross-sector and cross-border system boundaries.

In this work, we propose a CBA framework to assess this type of holistic ODP integrating EVs/ETs, RES, and the grid. The framework is designed and applied across three European countries over a 10-year horizon. It explicitly quantifies AI-driven operational benefits against the capital and operational expenditures of software, cloud infrastructure, and sensor networks.

While many platforms focus on specific functions — EV smart charging \citep{Schotetal2019} or RES forecasting \citep{Ahmedetal2020} — the ODP considered here takes a broader view. It coordinates charging-station availability, dynamic EV/ET fleet routing with energy and battery-health constraints, demand response programmes, and renewable generation forecasting within a single intelligent ecosystem. This integrated management boosts grid flexibility by aligning EV/ET charging with RES availability, enhances system reliability through predictive congestion management, and yields economic benefits via optimized procurement and trading — while potentially extending EV/ET battery lifespans. To quantify these outcomes, we apply a seven-step CBA framework informed first by the European Commission smart-grid guidance \citep{ECJRC2012} and then by the Isernia pilot application \citep{Stromsather2018}. 

Although the framework is grounded in established EU guidance \citep{ECJRC2012} and in a smart-grid pilot application \citep{Stromsather2018}, its contribution extends beyond application. To the authors' knowledge, prior CBA methodologies have not been tailored to operational digital platforms, where AI model development, cloud infrastructure, and continuous retraining pipelines constitute significant and distinct investment categories.

A key design choice concerns the mobility dimension. The proposed framework does not optimize EV/ET logistics directly. Accordingly, routing is included here as a conceptual ODP capability at the architectural level, whereas the case-study monetization captures routing-related value only through conservative proxy parameters within FES rather than through solved vehicle-routing instances. Instead, it quantifies the incremental system-level benefits of the ODP by parameterizing EV/ET characteristics — higher energy consumption per kilometre and reduced charging flexibility under duty schedules — alongside RES and grid interactions. These parameters feed into the benefit and cost channels (curtailment reduction, peak demand management, battery degradation savings) without requiring a full routing or scheduling model.

Benefits are defined as incremental relative to a \textit{without-ODP} baseline. Costs and benefits are linked to the platform via measurable KPI deltas — curtailment avoided, peak load reduced, imbalance costs lowered, congestion mitigated. Environmental valuations use country-specific, time-resolved emission factors rather than static averages. Uncertainty analysis covers AI forecast performance, adoption trajectories, and valuation parameters. Together, these design choices yield a methodological advance beyond single-infrastructure CBAs and provide a replicable model for appraising digital platforms that coordinate mobility, energy, and grid systems. The framework is illustrated through deployment in Austria, Hungary, and Slovenia, which provide a diverse Central European testbed with contrasting electrification trajectories, renewable build-out patterns, and market conditions, but should not be interpreted as statistically representative of all EU member states. Therefore, the main contributions of the paper are threefold:
\begin{enumerate}
\item [\textit{(i)}] Proposing a seven-step CBA framework to evaluate an AI-driven cross-sector, and cross-border ODP, explicitly identifying and monetizing AI-specific cost drivers (CAPEX and OPEX) and quantifying AI-driven benefit improvements (such as short-term RES forecasting and EV/ET charging optimization) as they propagate through the integrated transport–energy system. Performance improvements (e.g., forecast-error reduction, scheduling gains) are calibrated to literature benchmarks rather than outputs of an implemented ODP instance; the present study therefore constitutes an ex-ante assessment for the proposed ODP.
    
\item [\textit{(ii)}] Demonstrating a multi-layered, AI-driven ODP architecture by enumerating its core AI modules for RES forecasting, battery-health–aware EV charging, electric vehicle routing problem with time windows (E-VRPTW) routing for electric trucks, and co-optimization of grid ancillary services, and showing how verifiable performance metrics directly map to systemic economic value creation, thereby quantifying the  economic, reliability, and environmental benefits, and the interconnectedness of each layer in its architectural core.
 \item [\textit{(iii)}] Applying the proposed framework to country-based ex-ante case studies in Austria, Hungary, and Slovenia using country-specific input data and projections (CAPEX, OPEX, RES capacity factors, EV/ET fleet sizes and their projections, and wholesale price time series) and computing present-value streams for each benefit alongside CAPEX/OPEX. This provides actionable policy and managerial insights for Fit for 55 and REPowerEU implementation.
\end{enumerate}
The remainder of the paper is structured as follows: Section \ref{sec_related_work} reviews related work. Section \ref{meth} presents the AI-driven ODP architecture and the proposed CBA framework. Section \ref{application} presents the CBA steps with a case study and application. Section \ref{dis} discusses the main results of the study and some of the limitations. Finally, Section \ref{conc} offers conclusions and future directions.

\section{Related work}\label{sec_related_work}

This section contextualizes the proposed framework through a structured
literature review. AI-enabled coordination of mobility assets and grid operations,
RES integration and smart-grid management, and economic quantification of digital
energy systems are explored. We highlight how the proposed ODP addresses persistent
gaps in existing literature by functioning as an active, interoperable operational
layer. Moreover, we quantify the costs and benefits associated with these platforms
to advance the energy--mobility nexus.

\subsection{AI-enabled coordination of EV/ET charging, flexibility, and grid constraints}

A major stream of related work examines how electrified mobility affects power-system planning and operation. Large-scale EV/PHEV integration creates economic dispatch and risk-management challenges that motivate coordinated scheduling under uncertainty \citep{Peng2012PHEVEconomicDispatch}. Fast charging amplifies these pressures through load spikes and network congestion, requiring controlled charging strategies and regulatory preparedness \citep{Shariff2022FastCharging}. At the system level, flexibility requirements become more stringent as VRE penetration rises: \cite{Kaushik2022Flexibility} shows that beyond approximately 80\% VRE share, long-duration flexibility resources become critical for reliable operation.

At the distribution-network level, voltage and thermal constraints are repeatedly identified as binding factors under clustered charging. A smart-meter-based LV case study finds present-day EV penetration below 1\% per feeder, suggesting limited immediate stress but highlighting the need for forward-looking planning tools \citep{Rahman2022LowVoltageEVImpact}. The rapid growth of this research area is itself instructive: Ref. \cite{AriasLondono2020EVReview} traces EV--distribution-system interactions from 1973 to 2019 and notes an IEEE Xplore corpus of approximately $6.4\times10^4$ items, motivating structured taxonomies and decision-support frameworks.

From an AI-method perspective, \cite{Arevalo2024} conducts a PRISMA-based review of AI integration in EV energy management systems, identifying 46 highly relevant studies. Contributions cluster around forecasting, optimization/control, route planning, and predictive maintenance, predominantly using machine learning and metaheuristics. These findings support the ODP design choice to integrate sequence models for forecasting with mathematical optimization for scheduling within a unified data-and-control architecture. Critically, the review also reveals a persistent structural gap: most contributions optimize a single subsystem — vehicle, charger, or feeder — rather than orchestrating platform-level coordination across mobility, RES, and the grid simultaneously. Closing this end-to-end gap is the central architectural proposition of this paper. 

\subsection{AI for RES integration and smart-grid operation}

A second stream addresses RES variability and the need for anticipatory control. Ref. \cite{Nasim2024} surveys multiple AI strategy classes for renewable generation, forecasting, and optimization, finding that the field has matured from isolated predictors toward integrated decision pipelines. \cite{Liu2022} similarly emphasizes that high RES shares in multi-energy systems demand sophisticated forecasting and adaptive control, while flagging persistent challenges around non-stationarity and data availability. Smart-grid transformation work frames AI as a core capability for real-time analytics, predictive maintenance, demand response, and resilience \citep{Rajaperumal2025}. Related cyber-physical analyses consistently identify cybersecurity vulnerabilities as a first-order deployment risk \citep{Diaba2024CyberPhysical}. These insights motivate an architecture that explicitly forecasts both demand and RES generation, translates those forecasts into jointly feasible schedules under hard grid constraints, and treats practical deployment requirements — interoperability standards, cybersecurity controls, and scalable cloud infrastructure — as first-class architectural components rather than afterthoughts. 

\subsection{Cost--benefit analysis (CBA) of digital/AI-enabled energy systems}

While many studies report technical benefits of smart charging, V2G, and RES-aware scheduling, fewer works quantify the full economic and societal implications in a structured and transparent manner. A close methodological precedent is the CBA performed for the InteGRIDy project. Ref. \cite{Gudlaugsson2023} evaluates two pilots and reports payback periods of 8.2 years (Barcelona) and 2.8 years (St.~Jean), with a 22\% revenue increase observed for one pilot stakeholder. These results confirm that digitalized RES integration can be economically viable. However, they also expose a key limitation: CBAs are typically performed for single projects or sites and do not quantify the incremental value of \emph{integrated, cross-sector} AI functionalities spanning mobility, RES, and grid in a unified platform. Furthermore, conventional CBAs rarely disaggregate the CAPEX and OPEX associated with cloud infrastructure, AI model retraining pipelines, sensor networks, and real-time data ingestion — cost categories that are material for a holistic ODP and that directly condition whether projected KPI improvements are realized in practice.

The literature establishes well-documented technical challenges from EV/ET charging and fast charging, the importance of flexibility under high RES shares, and the promise of AI for forecasting and operational optimization. However, three gaps recur across the reviewed works:
 \textit{(i) Platform gap:} many contributions remain siloed (EV charging, RES forecasting, or grid optimization), whereas real deployments require a holistic platform that orchestrates these layers end-to-end.
     \textit{(ii) Operationalization gap:} practical issues—such as database ingestion, interoperability, scalability, and cybersecurity—are recognized but often not integrated into the proposed control/optimization workflow.
 \textit{(iii) Quantification gap:} economic assessments are frequently qualitative or case-specific, with limited emphasis on the \emph{incremental} costs and benefits attributable to integrated AI functionalities.

To address these gaps, this paper proposes an AI-driven ODP architecture for cross-sector (EVs/ETs--RES--grid) operational coordination, evaluated through a structured seven-step CBA framework. The explicit aim is to quantify incremental costs and benefits that are commonly treated qualitatively or with narrower scope in prior work. Table~\ref{tab:rw_matrix} summarizes how this work addresses the limitations of preceding literature.

\begin{table*}[htbp]
\centering
\caption{Summary of preceding works, limitations, and what this paper addresses. The proposed work explicitly unifies cross-sector operations while quantifying the outcomes via a structured CBA.
\cmark = explicitly covered; \pmark = partially/implicitly covered; \xmark = not covered.}
\label{tab:rw_matrix}
\small
\setlength{\tabcolsep}{6pt} 
\renewcommand{\arraystretch}{1.15}

\begin{tabular*}{\textwidth}{@{\extracolsep{\fill}}l c c c c c c@{}}
\toprule
Ref. 
& EV/ET
& RES
& Grid constr.
& ODP / interop.
& Cyber
& CBA / quant. \\
\midrule
Peng et al. \citep{Peng2012PHEVEconomicDispatch}     & \cmark & \xmark & \cmark & \xmark & \xmark & \pmark \\
Shariff et al. \citep{Shariff2022FastCharging}      & \cmark & \xmark & \cmark & \xmark & \xmark & \xmark \\
Kaushik et al. \citep{Kaushik2022Flexibility}       & \pmark & \cmark & \cmark & \xmark & \xmark & \xmark \\
Ar\'evalo et al. \citep{Arevalo2024}                & \cmark & \xmark & \pmark & \xmark & \xmark & \xmark \\
Rahman et al. \citep{Rahman2022LowVoltageEVImpact}  & \cmark & \xmark & \cmark & \xmark & \xmark & \xmark \\
Arias-Londo\~no et al. \citep{AriasLondono2020EVReview} & \cmark & \pmark & \cmark & \xmark & \xmark & \xmark \\
Wang et al. \citep{Wang2025}                        & \pmark & \cmark & \pmark & \pmark & \pmark & \xmark \\
Al Nasim et al. \citep{Nasim2024}                   & \xmark & \cmark & \xmark & \xmark & \xmark & \xmark \\
Liu et al. \citep{Liu2022}                          & \pmark & \cmark & \pmark & \xmark & \xmark & \xmark \\
Rajaperumal and Columbus \citep{Rajaperumal2025}    & \xmark & \pmark & \cmark & \pmark & \cmark & \xmark \\
Diaba et al. \citep{Diaba2024CyberPhysical}         & \xmark & \xmark & \pmark & \pmark & \cmark & \xmark \\
Gudlaugsson et al. \citep{Gudlaugsson2023}          & \xmark & \cmark & \pmark & \pmark & \xmark & \cmark \\
\midrule
The proposed framework    & \cmark & \cmark & \cmark & \cmark & \cmark & \cmark \\
\bottomrule
\end{tabular*}
\end{table*}

\section{AI-driven ODP and the CBA framework}\label{meth}
This section presents the AI-driven ODP system architecture and the proposed CBA framework. Section~\ref{sec_odp_architecture} describes the multi-layered ODP architecture and its AI engine core. Section~\ref{framework} introduces the structured seven-step CBA framework used to assess the financial and economic feasibility of the investment. 

\subsection{ODP system architecture and AI engine}\label{sec_odp_architecture}

The AI-driven ODP is an operational platform that integrates actors, technologies, and data across the energy and transport sectors. Table~\ref{tab_architecture} presents the functional architecture. The five layers—\textit{Perception, Data, Middleware, Business,} and \textit{Application}—organize how data are acquired, curated, exchanged, optimized, and communicated to users, thereby supporting modularity, scalability, and maintainability. AI functionality is supported by continuous monitoring, periodic retraining, and scalable computing resources. At the system level, the platform performs real-time data acquisition and validation, forecasts RES generation, load, and electricity prices, and supports EV/ET smart charging, V2X dispatch, demand response, grid balancing, fleet routing, and market settlement. It also provides monitoring, visualization, KPI reporting, cybersecurity, and data governance. These functions rely on internal databases for telemetry, operational records, and historical analytics, together with external inputs such as weather and NWP data, electricity prices, grid constraint data, charging-network status, and GIS and traffic information. In this context, the CBA evaluates the economic justification for deploying the functional capabilities of the ODP rather than any specific software-vendor stack.

\begin{footnotesize}
\begin{longtable}{>{\raggedright\arraybackslash}p{0.10\textwidth} >{\raggedright\arraybackslash}p{0.20\textwidth} >{\raggedright\arraybackslash}p{0.20\textwidth} >{\raggedright\arraybackslash}p{0.36\textwidth}}
\caption{Functional architecture, requirements, and representative implementation categories for the ODP integrating EV, ET, RES, and the grid.} \label{tab_architecture} \\
\toprule
\textbf{Layer} & \textbf{Requirement} & \textbf{Functional component} & \textbf{Representative implementation requirements} \\
\midrule
\endfirsthead

\multicolumn{4}{l}%
{{\bfseries Table~\thetable{} -- continued from previous page}} \\
\toprule
\textbf{Layer} & \textbf{Requirement} & \textbf{Functional component} & \textbf{Representative implementation requirements} \\
\midrule
\endhead

\midrule \multicolumn{4}{r}{{Continued on next page}} \\
\endfoot

\bottomrule
\endlastfoot

\rowcolor{perception}
\textbf{Perception layer} & Collect data on vehicles & EV/ET sensors & BMS sensors, GPS, OCPP-compliant charge monitors \\
\rowcolor{perception}
 & Collect data on RES & RES sensors & Pyranometers, anemometers, power meters \\
\rowcolor{perception}
 & Collect data on environmental conditions & Environmental sensor arrays & IoT weather nodes, multi-sensor modules \\
\rowcolor{perception}
 & Collect data on weather conditions & Weather sensors & Integrated weather stations, satellite APIs \\
\rowcolor{perception}
 & Transmit data & Connectivity solutions & 5G, LTE-M/NB-IoT, LoRaWAN, Satellite \\
\midrule
\rowcolor{data}
\textbf{Data layer} & Receive data & Communication protocols & MQTT, HTTP/HTTPS, OPC-UA, AMQP \\
\rowcolor{data}
 & Communicate data & Message queues & Apache Kafka, RabbitMQ \\
\rowcolor{data}
 & Store data & Databases & PostgreSQL, MongoDB, InfluxDB, TimescaleDB \\
\rowcolor{data}
 & Store big data & Cloud/object storage and data lakes & Scalable object storage, archival layers, batch analytics support \\
\rowcolor{data}
 & Move data & Data integration and ETL & Stream/batch ingestion, data transformation, orchestration \\
\midrule
\rowcolor{middleware}
\textbf{Middleware layer} & Integrate with existing systems & Context broker and integration middleware & Service orchestration, enterprise integration, interoperability adapters \\
\rowcolor{middleware}
 & Provide clean APIs & API gateway and identity management & Authentication, rate limiting, and secure API exposure \\
\rowcolor{middleware}
 & Integrate with EU data spaces & Data-space interoperability connectors & Federated data exchange, trust, and policy enforcement \\
\rowcolor{middleware}
 & Collect market data & Market and operator data interfaces & Wholesale-market, balancing, and system-operator data feeds \\
\midrule
\rowcolor{business}
\textbf{Business layer (AI Engine Core)} & Predict grid load, RES, prices & Forecasting engine & Sequence models with NWP features; scalable training and inference \\
\rowcolor{business}
 & Optimize EV/ET charging, V2X/V2B & Optimization and real-time control & MILP/MIQP scheduling, RL agents, battery models \\
\rowcolor{business}
 & Manage grid flexibility & Demand-response orchestration & OpenADR-compliant dispatch, baselining, and load aggregation \\
\rowcolor{business}
 & Optimize EV/ET routing & Fleet routing and dispatch engine & Electric VRP solvers, charger-aware routing, metaheuristics \\
\rowcolor{business}
 & Weather forecast integration & Weather-data assimilation and bias correction & NWP feeds, AI correction, sensor fusion \\
\midrule
\rowcolor{application}
\textbf{Application layer} & User interaction & Mobile/web user interfaces & Charge planning, fleet dashboards, operator views \\
\rowcolor{application}
 & Navigation to charging points & Navigation and charger-location services & Route guidance, charger availability, map integration \\
\rowcolor{application}
 & ODP insights & Analytics and visualization & KPI dashboards, reporting, and monitoring \\
\rowcolor{application}
 & Administration tools & Administration and model-governance tools & Drift detection, admin controls, audit logs \\
\end{longtable}
 \begin{tablenotes}
\item Note: The ODP architecture is based on the EU-funded BEGONIA project. Table~\ref{tab_architecture} reports functional categories, while detailed example technology mappings are documented in Deliverable D3.1 \cite{begonia_d3.1_2024}.
 \end{tablenotes}
\end{footnotesize}

Note that the scope of this section is the economic and financial appraisal of the ODP; we do not empirically validate or optimize the AI modules. Performance figures (e.g., forecasting accuracy, coordination gains) are drawn from the literature as calibrated benchmarks rather than measured case-study outputs. Each architectural layer description below should be read together with the technological diagram in Fig.~\ref{Schematic}, which illustrates platform-level data flows. The CBA in Section~\ref{application} then evaluates the economic and financial implications of implementing this ODP across the three case-study countries.

The \textit{Perception layer} acquires real-time telemetry from EVs, ETs, RES assets, charging infrastructure, and environmental sensors at granular resolutions (e.g., 1-minute intervals), transmitting data over 5G, LTE-M, or LoRaWAN links.

The \textit{Data layer} ingests, validates, and stores these heterogeneous streams in a hybrid architecture of time-series databases and data lakes, with integrated anomaly detection and missing-data imputation to guarantee robust AI inputs.

The \textit{Middleware layer} provides interoperability with utility SCADA/EMS systems, wholesale-market and system-operator interfaces, and EU data spaces, ensuring secure, standardized, cross-sectoral data flow.

The \textit{Business layer} constitutes the AI engine core. Its four modules---and their CBA relevance---are as follows: (i)~\textit{Forecasting}: sequence models predict RES generation, grid load, and electricity prices at 15-min to 72-h horizons, providing the input signals on which all downstream optimization depends. (ii)~\textit{Smart charging \& V2X dispatch}: mathematical-programming models schedule individual and fleet-level charging against RES availability, tariffs, and grid limits, while reinforcement-learning agents handle real-time V2G/V2B dispatch with battery-health safeguards as hard constraints. The CBA captures the resulting KPI deltas---reduced peak demand, avoided curtailment, and lower imbalance costs---relative to a \textit{without-ODP} baseline. (iii)~\textit{Fleet routing}: electric-vehicle-routing solvers jointly minimize energy, time, and degradation costs under real-time traffic, weather, and charger-availability constraints. (iv)~\textit{Demand response}: standards-compliant orchestration tools aggregate flexible EV/ET loads and manage ancillary-service participation using AI-driven baselining. These models operate within a monitored MLOps workflow with scheduled retraining to counter data and concept drift.

Finally, the \textit{Application layer} delivers user-facing mobile and web applications (charge planning, fleet dashboards, route guidance) and operator consoles with KPI analytics, providing transparent visibility into AI-driven decisions across the platform.

Taken together, the five layers operate as a closed-loop operational pipeline rather than as a catalogue of standalone modules. Sensor and market data move upward through ingestion, validation, and interoperability services; the AI engine converts those standardized states into forecasts, schedules, and dispatch decisions; and the application layer returns decisions, alerts, and KPIs to users and operators, whose realized outcomes feed back into model monitoring and retraining. This end-to-end coupling is the key architectural distinction that enables the CBA to attribute monetized value to specific platform functions instead of treating digitalization as a generic uplift factor.

\begin{figure}[htbp]
\centering
\resizebox{\textwidth}{!}{%
\begin{tikzpicture}[
  >=Stealth,
  lhdr/.style={
    font=\bfseries\fontsize{8.5}{10}\selectfont,
    text=white, fill=#1, rounded corners=2pt,
    inner sep=4pt, anchor=west},
  box/.style 2 args={
    draw=#1, fill=#2,
    rounded corners=3pt,
    minimum height=0.72cm,
    font=\fontsize{7.5}{9}\selectfont,
    text=black, align=center,
    line width=0.5pt,
    inner xsep=5pt, inner ysep=3pt},
  ebox/.style={box={gray!70}{gray!10}},
  uflow/.style={
    -{Stealth[length=2.2mm,width=1.8mm]},
    line width=0.8pt, draw=#1!80},
  biflow/.style={
    {Stealth[length=2mm,width=1.6mm]}-{Stealth[length=2mm,width=1.6mm]},
    line width=0.7pt, draw=#1!80},
  feedbk/.style={
    {Stealth[length=1.8mm,width=1.5mm]}-{Stealth[length=1.8mm,width=1.5mm]},
    line width=0.5pt, draw=#1!80, densely dashed},
  conn/.style={
    draw=#1!50, line width=0.5pt},
  flbl/.style={
    font=\fontsize{5.5}{7}\selectfont, text=darkgray,
    fill=white, inner sep=1.5pt, rounded corners=0.5pt},
]


\fill[cyan!10, rounded corners=5pt, draw=cyan!40, line width=0.5pt] (0,0) rectangle (13.0, 1.4);
\node[lhdr=cyan!80!black] at (0.15, 1.15) {Perception layer};

\fill[blue!10, rounded corners=5pt, draw=blue!40, line width=0.5pt] (0, 1.8) rectangle (13.0, 3.2);
\node[lhdr=blue!80!black] at (0.15, 2.95) {Data layer};

\fill[teal!10, rounded corners=5pt, draw=teal!40, line width=0.5pt] (0, 3.6) rectangle (13.0, 5.0);
\node[lhdr=teal!80!black] at (0.15, 4.75) {Middleware layer};

\fill[violet!10, rounded corners=5pt, draw=violet!40, line width=0.5pt] (0, 5.4) rectangle (13.0, 8.0);
\node[lhdr=violet!80!black] at (0.15, 7.75) {Business layer (AI engine core)};

\fill[purple!10, rounded corners=5pt, draw=purple!40, line width=0.5pt] (0, 8.4) rectangle (13.0, 9.8);
\node[lhdr=purple!80!black] at (0.15, 9.55) {Application layer};


\node[box={cyan}{white}, minimum width=2.0cm] (pEV)   at (1.5, 0.5) {EV/ET sensors};
\node[box={cyan}{white}, minimum width=2.0cm] (pRES)  at (4.0, 0.5) {RES sensors};
\node[box={cyan}{white}, minimum width=2.2cm] (pGrid) at (6.5, 0.5) {Grid \& weather\\[-1pt]sensors};
\node[box={cyan}{white}, minimum width=2.2cm] (pCP)   at (9.2, 0.5) {Charging infra.\\[-1pt]monitors};
\node[box={cyan}{white}, minimum width=1.6cm] (pComm) at (11.7, 0.5) {5G / IoT\\[-1pt]comms};

\node[box={blue}{white}, minimum width=2.4cm] (dIng)   at (2.5, 2.3) {Data ingestion\\[-1pt]\& validation};
\node[box={blue}{white}, minimum width=2.4cm] (dStore) at (6.5, 2.3) {Storage (DBs,\\[-1pt]Data Lakes)};
\node[box={blue}{white}, minimum width=2.4cm] (dMsg)   at (10.5, 2.3) {Message brokers\\[-1pt]\& ETL pipelines};

\node[box={teal}{white}, minimum width=2.4cm] (mInt) at (2.5, 4.1) {Integration\\[-1pt]Middleware};
\node[box={teal}{white}, minimum width=2.4cm] (mAPI) at (6.5, 4.1) {API Gateway\\[-1pt]\& Security};
\node[box={teal}{white}, minimum width=2.4cm] (mDS)  at (10.5, 4.1) {Data Space\\[-1pt]Connectors};

\node[box={violet}{white}, minimum width=2.3cm, minimum height=0.85cm] (aFore)  at (1.5, 7.1) {RES / Load / Price\\[-1pt]forecasting\\[-1pt]{\fontsize{5.5}{6.5}\selectfont\itshape LSTM, attention}};
\node[box={violet}{white}, minimum width=2.3cm, minimum height=0.85cm] (aChg)   at (4.0, 7.1) {Smart charging\\[-1pt]\& V2X dispatch\\[-1pt]{\fontsize{5.5}{6.5}\selectfont\itshape MILP, RL}};
\node[box={violet}{white}, minimum width=2.3cm, minimum height=0.85cm] (aRoute) at (6.5, 7.1) {ET routing \&\\[-1pt]logistics\\[-1pt]{\fontsize{5.5}{6.5}\selectfont\itshape E-VRPTW}};
\node[box={violet}{white}, minimum width=2.5cm, minimum height=0.85cm] (aDR)    at (9.2, 7.1) {Demand response\\[-1pt]\& grid balancing\\[-1pt]{\fontsize{5.5}{6.5}\selectfont\itshape OpenADR}};
\node[box={violet}{white}, minimum width=2.0cm, minimum height=0.85cm] (aWeath) at (11.7, 7.1) {Weather\\[-1pt]modeling\\[-1pt]{\fontsize{5.5}{6.5}\selectfont\itshape NWP APIs}};

\node[box={violet}{white}, minimum width=1.9cm] (sPric) at (2.75, 6.0) {Pricing \&\\[-1pt]Flexibility};
\node[box={violet}{white}, minimum width=2.0cm] (sChgC) at (6.5, 6.0) {Charging\\[-1pt]coordination};
\node[box={violet}{white}, minimum width=2.0cm] (sGrid) at (10.45, 6.0) {Grid ancillary\\[-1pt]\& Settlement};

\draw[conn=violet] (aFore.south) -- (sPric.north);
\draw[conn=violet] (aChg.south)  -- (sPric.north);
\draw[conn=violet] (aChg.south)  -- (sChgC.north);
\draw[conn=violet] (aRoute.south)-- (sChgC.north);
\draw[conn=violet] (aDR.south)   -- (sGrid.north);
\draw[conn=violet] (aWeath.south)-- (sGrid.north);

\node[box={purple}{white}, minimum width=2.0cm] (aMob)   at (1.5, 9.0) {Driver mobile\\[-1pt]apps \& nav.};
\node[box={purple}{white}, minimum width=2.0cm] (aFleet) at (4.0, 9.0) {Fleet mgmt.\\[-1pt]dashboards};
\node[box={purple}{white}, minimum width=2.0cm] (aKPI)   at (6.5, 9.0) {KPI analytics\\[-1pt]\& visualization};
\node[box={purple}{white}, minimum width=2.0cm] (aOps)   at (9.2, 9.0) {Grid operator\\[-1pt]consoles};
\node[box={purple}{white}, minimum width=1.6cm] (aAdmin) at (11.7, 9.0) {Admin UI\\[-1pt]\& rules};


\draw[uflow=cyan] (pEV.north)   -- (1.5, 1.8);
\draw[uflow=cyan] (pRES.north)  -- (4.0, 1.8);
\draw[uflow=cyan] (pGrid.north) -- (6.5, 1.8);
\draw[uflow=cyan] (pCP.north)   -- (9.2, 1.8);
\draw[uflow=cyan] (pComm.north) -- (11.7, 1.8);
\node[flbl] at (6.5, 1.6) {telemetry, status, generation data};

\draw[uflow=blue] (dIng.north)   -- (mInt.south);
\draw[uflow=blue] (dStore.north) -- (mAPI.south);
\draw[uflow=blue] (dMsg.north)   -- (mDS.south);
\node[flbl] at (6.5, 3.4) {cleaned data, historical archives, streams};

\draw[uflow=teal] (mInt.north) -- (sPric.south);
\draw[uflow=teal] (mAPI.north) -- (sChgC.south);
\draw[uflow=teal] (mDS.north)  -- (sGrid.south);
\node[flbl] at (6.5, 5.2) {API events, market signals, formatted states};

\draw[uflow=violet] (aFore.north) -- (aMob.south);
\draw[uflow=violet] (aChg.north)  -- (aFleet.south);
\draw[uflow=violet] (aRoute.north)-- (aKPI.south);
\draw[uflow=violet] (aDR.north)   -- (aOps.south);
\draw[uflow=violet] (aWeath.north)-- (aAdmin.south);
\node[flbl] at (6.5, 8.2) {schedules, KPIs, alerts, dispatch signals};

\draw[feedbk=purple] (aAdmin.south) -- (aDR.north) node[flbl, pos=0.4, right=2pt] {\itshape prefs};
\draw[feedbk=violet] (sGrid.south) -- (mDS.north) node[flbl, pos=0.5, right=2pt] {\itshape control commands};

\node[ebox, minimum width=2.1cm] (xMkt) at (15.5, 7.5) {Energy markets\\[-1pt]{\fontsize{5}{6}\selectfont EPEX, Nord Pool}};
\node[ebox, minimum width=2.1cm] (xTSO) at (15.5, 6.0) {TSO / DSO\\[-1pt]{\fontsize{5}{6}\selectfont SCADA, EMS}};
\node[ebox, minimum width=2.1cm] (xDS)  at (15.5, 4.1) {EU data spaces\\[-1pt]{\fontsize{5}{6}\selectfont Gaia-X, ENTSO-E}};
\node[ebox, minimum width=2.1cm] (xWth) at (15.5, 2.3) {Weather \&\\[-1pt]{\fontsize{5}{6}\selectfont traffic APIs}};

\node[draw=gray!50, rounded corners=4pt, line width=0.5pt,
      fit=(xMkt)(xTSO)(xDS)(xWth),
      inner xsep=5pt, inner ysep=6pt,
      label={[font=\bfseries\fontsize{7.5}{9}\selectfont,
              text=darkgray, anchor=south]north:{External Systems}}] {};

\draw[biflow=gray] (aDR.east) -- ++(0.8,0) |- (xMkt.west);
\node[flbl] at (13.8, 7.5) {bids, prices};
\draw[biflow=gray] (aWeath.east) -- ++(0.3,0) |- (xTSO.west);
\node[flbl] at (13.8, 6.0) {grid limits};
\draw[biflow=gray] (mDS.east) -- (xDS.west);
\node[flbl] at (13.8, 4.1) {cross-border};
\draw[biflow=gray] (dMsg.east) -- (xWth.west);
\node[flbl] at (13.8, 2.3) {weather data};

\fill[brown!10, rounded corners=3pt, draw=brown!50, line width=0.4pt]
  (-0.75, 0) rectangle (-0.15, 9.8);
\node[rotate=90, font=\bfseries\fontsize{7}{9}\selectfont, text=brown!80!black]
  at (-0.45, 4.9) {OPTIONAL SECURE LOG};

\foreach \y in {0.7, 2.5, 4.3, 6.7, 9.1}{
  \draw[brown!60, line width=0.35pt, -{Stealth[length=1.2mm]}]
    (-0.15,\y) -- (0.05,\y);
}

\draw[decorate, decoration={brace, amplitude=4pt, mirror},
      gray, line width=0.4pt]
  (-1.05, 0) -- (-1.05, 9.8)
  node[midway, left=5pt, rotate=90,
       font=\fontsize{5.5}{7}\selectfont\itshape, text=darkgray]
    {End-to-end real-time data pipeline};

\end{tikzpicture}
}
\caption{Technological diagram of the core ODP architecture mapping to its five principal layers. Data flows bottom-to-top across five layers. The Perception layer collects telemetry from EV/ET, RES, grid, and weather sensors. The Data layer ingests, validates, and stores these streams. The Middleware layer provides standardized interoperability via API gateways and data-space connectors. The Business layer runs the AI engine: forecasting, smart-charging optimization, fleet routing, and demand-response dispatch. The Application layer exposes outputs through dashboards and mobile apps. Bidirectional arrows (right) show data exchange with external systems (energy markets, TSOs/DSOs, EU data spaces). An optional secure-logging sidebar indicates a transaction-record layer that is not modeled as a separate value driver in the CBA.}
\label{Schematic}
\end{figure}

Fig.~\ref{Schematic} illustrates how the same five layers interact. The secure-logging layer shown in Fig.~\ref{Schematic} is optional and, where implemented, its costs are subsumed within general platform-infrastructure and cybersecurity expenditure rather than treated as a separate monetized benefit stream. 

The ODP--CBA relies on country-specific inputs: RES generation forecasts, EV/ET fleet profiles and charging schedules, grid constraint data, and cross-border flow parameters. These data are increasingly standardized across Europe through smart-grid deployments, national data hubs, and harmonized API interfaces. The framework is designed to connect directly to these interoperable data rails, making large-scale deployment practical in any jurisdiction where electrification and data-sharing initiatives are underway.

\subsection{Cost-benefit analysis framework}\label{framework}
 
To evaluate the proposed AI-driven ODP, we employ a structured seven-step CBA framework inspired by the European Commission smart-grid guidelines \citep{ECJRC2012} and the Isernia pilot study \citep{Stromsather2018}. The framework captures AI-related costs --- platform development, sensors and IoT devices, model training and maintenance --- alongside the benefits directly attributable to AI-driven performance improvements. The seven steps proceed as follows: (1) precise project definition covering objectives, engineering features, geographic scope, and stakeholders; (2) mapping enabling technologies to operational platform functions; (3) linking those functions to economic, reliability, and environmental benefits; (4) quantifying and monetizing each benefit using transparent formulas; (5) tallying CAPEX and OPEX over the project lifecycle; (6) comparing costs and benefits via NPV and BCR; and (7) conducting sensitivity analysis and Monte Carlo simulation. Table~\ref{tab_seven-step-cba} summarizes these steps.
\begin{table}[ht]
  \centering
  \begin{threeparttable}
  \caption{Seven-step framework for CBA, applied to an AI-driven cross-sectoral platform.}
  \label{tab_seven-step-cba}
  \begin{tabular}{@{}p{0.08\linewidth}p{0.9\linewidth}@{}}
    \toprule
    \textbf{Step} & \textbf{Procedure} \\
    \midrule
    1 & Define the project (scope, objectives, engineering features, AI functionalities, and their specific role). \\
    2 & Map technologies (including AI modules and their specific functionalities) to platform functions. \\
    3 & Map platform functions to economic, reliability, and environmental benefits. \\
    4 & Monetize and quantify benefits (linking ODP performance to benefit magnitude, with transparent assumptions). \\
    5 & Quantify costs (CAPEX and OPEX, including detailed AI system development, training, and maintenance costs). \\
    6 & Compare costs and benefits (NPV, BCR). \\
    7 & Conduct sensitivity analysis and risk assessment (including discount rate, technology adoption variability parameters, and Monte Carlo simulation). \\
    \bottomrule
  \end{tabular}
  \end{threeparttable}
\end{table}
Step~1 defines project scope, objectives, geographic coverage, stakeholders, and counterfactual baseline. Step~2 links each ODP technology layer and AI component to its operational function: data acquisition, system integration, forecasting and optimization, and decision support. Step~3 identifies how those functions generate economic, reliability, and environmental benefits. Step~4 converts each benefit into monetary terms using transparent formulas and country-specific data. Step~5 aggregates all CAPEX and OPEX for deployment and operation, including AI model development, sensors, cloud services, and personnel. Step~6 computes NPV and benefit--cost ratio (BCR) by discounting all cash flows at the social discount rate. Step~7 tests result robustness through scenario analysis and Monte Carlo simulation, and identifies key risks and mitigation strategies.

Conventional smart-grid CBAs typically treat ICT and analytics benefits qualitatively or aggregate them into a single ``smart-grid uplift'' factor. The ODP-tailored CBA proposed here differs in four structural respects: (i) It maps each AI module and platform function to specific, measurable KPI deltas (e.g., forecast error reduction leads to imbalance cost reduction; optimized charging leads to peak demand reduction and curtailment avoidance). (ii) It monetizes those deltas with transparent, reproducible formulas anchored to country-specific data. (iii) It includes AI-specific CAPEX and OPEX line items (model development, cloud inference, retraining pipelines, cybersecurity, data engineering) that are absent from conventional infrastructure CBAs. (iv) It explicitly tests AI performance uncertainty (forecast accuracy, optimization efficiency) and data availability alongside standard economic parameters in sensitivity and Monte Carlo analysis.

This architecture-to-benefit traceability ensures that the CBA is structurally linked to the ODP design. To improve reproducibility and technical transparency, Algorithm~1 formalizes the ODP-tailored CBA as a stepwise computational workflow with explicit baseline construction, KPI-to-benefit mapping, discounting, and uncertainty propagation.

\begin{algorithm}[H]
\caption{ODP-tailored CBA with uncertainty propagation: computational workflow linking platform architecture, KPI deltas, monetized benefit streams, and uncertainty-aware investment metrics.}
\label{alg_odp_cba}
\begin{algorithmic}[1]
\Require Country set $\mathcal{C}$; years $t\in\{2026,\dots,2035\}$; KPI set $\mathcal{K}$; benefit streams $\mathcal{B}$; cost items $\mathcal{J}$; discount rate $r$; scenario set $\mathcal{S}$; Monte Carlo samples $N$
\Ensure $\mathrm{NPV}$, $\mathrm{BCR}$, sensitivity ranking, and probabilistic distributions of outcomes
\State Construct baseline trajectories $X^{\mathrm{base}}_{c,t}$ (without ODP) and ODP trajectories $X^{\mathrm{ODP}}_{c,t}$ using harmonized country assumptions
\For{each country $c\in\mathcal{C}$ and year $t$}
    \For{each KPI $k\in\mathcal{K}$}
        \State Compute KPI delta: $\Delta k_{c,t}=k^{\mathrm{ODP}}_{c,t}-k^{\mathrm{base}}_{c,t}$
    \EndFor
    \For{each benefit stream $b\in\mathcal{B}$}
        \State Map KPI deltas to physical quantity: $Q_{b,c,t}=g_b\!\left(\{\Delta k_{c,t}\}_{k\in\mathcal{K}}\right)$
        \State Monetize stream: $B_{b,c,t}=Q_{b,c,t}\cdot P_{b,c,t}$
    \EndFor
    \For{each cost item $j\in\mathcal{J}$}
        \State Estimate annual cost: $C_{j,c,t}=C^{\mathrm{CAPEX}}_{j,c,t}+C^{\mathrm{OPEX}}_{j,c,t}$
    \EndFor
\EndFor
\State Aggregate: $B_t=\sum_{c}\sum_{b}B_{b,c,t}$ and $C_t=\sum_{c}\sum_{j}C_{j,c,t}$
\State Discount: $\mathrm{PV}(B)=\sum_t B_t/(1+r)^{t-2025}$;\quad $\mathrm{PV}(C)=\sum_t C_t/(1+r)^{t-2025}$
\State Compute: $\mathrm{NPV}=\mathrm{PV}(B)-\mathrm{PV}(C)$;\quad $\mathrm{BCR}=\mathrm{PV}(B)/\mathrm{PV}(C)$
\For{each deterministic scenario $s\in\mathcal{S}$}
    \State Recompute $(\mathrm{NPV}_s,\mathrm{BCR}_s)$ under perturbed assumptions (adoption rates, prices, discount rate, AI performance)
\EndFor
\For{$n=1$ \textbf{to} $N$}
    \State Sample uncertain inputs (forecast error, adoption, valuation factors, AI efficiency, data availability)
    \State Recompute $\mathrm{NPV}^{(n)}$ and $\mathrm{BCR}^{(n)}$
\EndFor
\State \Return point estimates, confidence intervals, and downside-risk probabilities (e.g., $\Pr[\mathrm{NPV}<0]$)
\end{algorithmic}
\end{algorithm}

\section{Cost-benefit analysis steps with case study}\label{application}

We now apply the seven-step CBA framework to the AI-driven ODP project across Austria (AT), Hungary (HU), and Slovenia (SI). The case-study design follows three principles: (i) heterogeneity across electrification pathways, (ii) comparability under a unified methodological baseline, and (iii) policy relevance for cross-border interoperability. Austria, Hungary, and Slovenia jointly satisfy these criteria by combining different RES expansion trajectories, EV/ET adoption levels, and market/grid characteristics while remaining operationally connected in the regional power context. Inputs are organized into three assumption blocks---country-specific, technology-specific, and macroeconomic---to preserve traceability of parameter effects. All input assumptions, input parameters, and sources are provided in Tables (\ref{tab_country_assumptions_revised}, \ref{eq_rap_appendix}, \ref{tab_params}, and \ref{tab_ai_costs}). For the economic appraisal, all future cash flows are expressed in constant 2025 Euros and discounted to present value using an economic discount rate of 4\% \citep{EUDiscountRateGuide2021}.
The project’s primary objective is to use real-time data and advanced AI models to optimize system operations. In doing so, it will enhance grid flexibility, lower energy procurement and operating costs, cut greenhouse-gas and local-air-pollutant emissions, improve power-system reliability, and enable seamless interoperability between transport electrification and grid operations.

\subsection*{Step 1: Define the project}\label{meth_step1}

The project is the deployment and operation of the AI-driven ODP for integrated management of EVs, ETs, RES, and the electric grid across three countries. Using real-time data and advanced AI models, the ODP pursues five operational objectives: enhanced grid flexibility, lower energy procurement and operating costs, reduced greenhouse-gas and local air-pollutant emissions, improved power-system reliability, and seamless interoperability between transport electrification and grid operations. The implementation scale is national/regional, with the three countries selected to provide a heterogeneous Central European testbed rather than a statistically representative sample of EU electrification contexts. 

Key engineering features of the ODP critical to the project include: \textit{(i)} AI-powered predictive algorithms for demand forecasting (EV/ET charging, overall grid load), RES generation forecasting, and energy price forecasting. \textit{(ii)} Optimization engines for smart charging scheduling, V2X services, electric fleet logistics (routing and charging), and demand response dispatch. These engines optimize multi-objective functions, including cost reduction, emissions reduction, and battery health preservation, thus enabling coordinated control strategies that balance the grid and maximize economic returns. \textit{(iii)} IoT infrastructure comprising a dense network of sensors and smart meters to provide granular, real-time monitoring of vehicles, charging stations, grid nodes, and RES installations. This high-resolution data collection is foundational for the AI engine’s situational awareness. \textit{(iv)} High-speed, reliable communication networks ensuring low-latency data transfer and control signal dissemination. This is critical for real-time AI inference and control, allowing the ODP to rapidly respond to changing conditions. \textit{(v)} Robust middleware integration with existing energy management systems (utility SCADA/DMS/EMS) and market platforms. By using standardized interfaces and data exchange protocols, the ODP can seamlessly interoperate with legacy systems and participate in energy markets, thus embedding the platform within the broader energy ecosystem.

\begin{figure}
\centering
\includegraphics[width=0.65\linewidth]{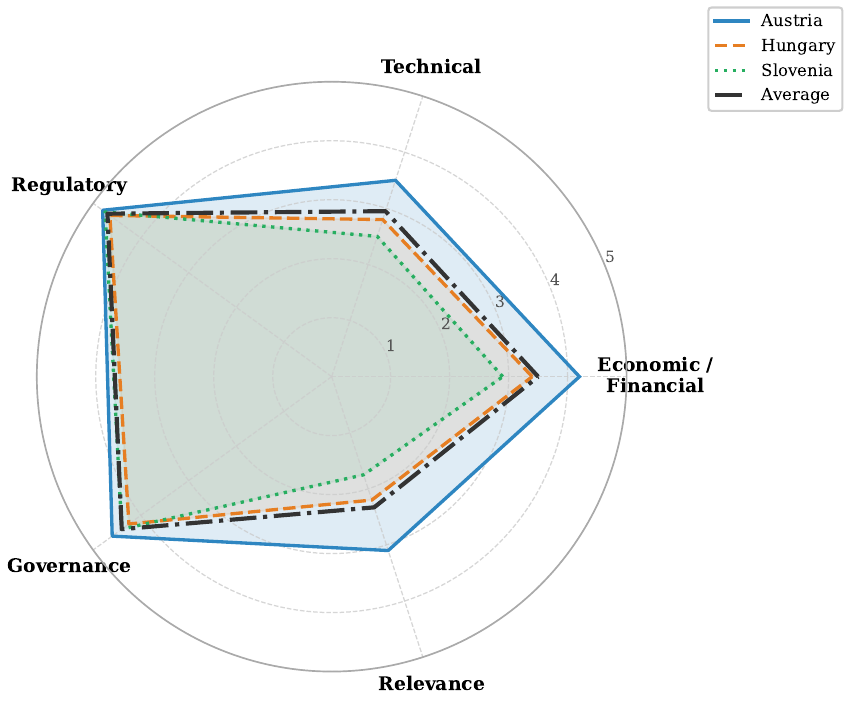}
\caption{Stakeholder-informed contextual assessment across regulatory, technical, governance, economic/financial, and relevance dimensions.}
\label{Radar}
\end{figure}

As part of the broader ODP design process, stakeholder input from TSOs/DSOs, CPOs, fleet operators, and regulatory stakeholders was used to identify functional requirements, data-sharing constraints, and implementation frictions relevant to the three-country setting. Figure~\ref{Radar} should therefore be interpreted as a stakeholder-informed contextual assessment of governance, relevance, technical readiness, regulatory fit, and economic/financial feasibility rather than as a standalone statistical survey result. These inputs are used to contextualize adoption assumptions and implementation risks in the CBA.

Fig. \ref{Radar} summarizes a stakeholder-informed assessment across regulatory, technical, governance, economic/financial, and relevance dimensions for the three case-study countries, providing contextual evidence on organizational readiness and adoption feasibility that complements the quantitative CBA.

The CBA benchmark is the \emph{without-ODP} system. The current baseline is characterized by largely uncoordinated EV/ET charging, minimal V2G capability, residual RES curtailment during surplus periods, and limited operator visibility due to legacy forecasting tools. All reported benefits are \emph{incremental} relative to this baseline. For benefit stream $k$ at time $t$,
\[\Delta B_{k,t} = \mathcal{M}_k(\text{ODP},t)-\mathcal{M}_k(\text{Baseline},t),\]
where $\mathcal{M}_k$ denotes the physical metric to be monetized, and $\Delta B_{k,t}$ denotes the annual monetized benefit for stream $k$ at time $t$ (aggregated across countries in reporting).
Participation by EV drivers and fleet operators in smart charging and demand response is modeled explicitly as a scenario input. The CBA assumes rational, benefit-maximizing behavior within country-specific policies and grid codes (EV/ET, RES, network operation). Non-technical barriers (heterogeneous charging habits, DR fatigue, data-sharing and regulatory frictions) are not modeled as separate processes; instead, they are reflected through conservative/central/optimistic participation ranges and assessed qualitatively in the stakeholder-informed multi-criteria evaluation (regulatory, governance, relevance; Fig. \ref{Radar}).

\subsection*{Step 2: Map technologies to functions}\label{meth_step2}
In this step, each core technology component of the ODP, particularly its AI modules as detailed in Section \ref{sec_odp_architecture}, is systematically mapped to the specific operational functions it enables. The ODP integrates a comprehensive suite of technologies, which together deliver several key functions for the system’s operation:\\
\textit{(A) Advanced data and information collection \& processing}: Real-time gathering, validation, and processing of granular data from EVs/ETs, RES installations, grid sensors, charging infrastructure status, and external sources (weather, market prices). This function is enabled by IoT sensors, connectivity solutions, and the ODP’s Data layer, ensuring that the AI engine has accurate and timely information about the system’s state at all times.\\
\textit{(B) Enhanced interoperability and system integration}: Seamless data exchange and coordinated operation among diverse systems and stakeholders (vehicles, grid operators, CPOs, energy markets, end-users). This is facilitated by the Middleware layer and use of standardized communication protocols and interfaces. Enhanced interoperability ensures that decisions made by the AI core can be effectively communicated and executed across different subsystems in real time.\\
 \textit{(C) AI-driven forecasting, optimization, and control}: This core function is where the ODP's AI engine directly drives value. It has two sub-functions:\\
 \textit{i) Intelligent forecasting}: The AI models (LSTMs, ML models, etc.,) provide highly accurate predictions for EV/ET charging demand, RES generation, grid load, and electricity prices. These forecasts improve upon traditional methods, reducing uncertainty and prediction error in the system’s operation.\\
  \textit{ii) Smart load management and grid balancing}: AI optimization algorithms enable dynamic smart charging/discharging of EVs/ETs, optimal dispatch of V2X services for grid support, dynamic EV/ET routing considering energy and battery health, and coordinated demand response actions. These control decisions continuously balance supply and demand, flatten load curves, alleviate grid constraints, and maximize the use of renewable energy.\\
\textit{(D) Decision support and automation}: User-friendly interfaces (mobile apps, dashboards) provide actionable insights and recommendations to end-users (e.g., optimal charging times) and system operators (e.g., congestion alerts and recommended mitigation actions). Automated control systems execute optimized schedules and dispatch signals directly generated by the AI engine, ensuring real-time responses and reducing manual intervention.

Table \ref{tab_mapping-tech} provides a high-level mapping of the ODP’s major technological assets and AI components to these functional categories A–D. For instance, IoT sensors and connectivity primarily support data collection (Function A). The integration middleware and communication protocols enable Function B. The Forecasting engine and optimization algorithms directly drive Function C (intelligent forecasting and control), while the user interfaces and automated control modules deliver Function D (decision support and action). This detailed mapping ensures that all critical functions necessary for the ODP's success are supported by specific, well-defined technological capabilities, with the core AI engine acting as the central intelligence that orchestrates these functions.

\begin{table*}[t!]
\centering
\caption{Mapping of ODP technologies and AI components to key functions: A = Data collection \& processing; B = System integration; C = AI forecasting, optimization \& control; D = Decision support \& automation.}
\label{tab_mapping-tech}
\begin{tabular}{@{}p{0.55\linewidth}cccc@{}}
\toprule
\textbf{Technology (asset) / AI component} & \textbf{A} & \textbf{B} & \textbf{C} & \textbf{D} \\
\midrule
IoT sensors (EV, ET, RES, weather, grid) & \checkmark &  &   &    \\
Edge gateways    & \checkmark & \checkmark &  &    \\
5G / LTE-M / LPWAN connectivity  & \checkmark & \checkmark &   &  \\
Message queue systems    & \checkmark & \checkmark &    & \\
Communication protocols (MQTT, OCPP, OpenADR)    & \checkmark & \checkmark &    & \\
Microservice architecture &  & \checkmark & \checkmark & \\
Data integration middleware &  & \checkmark &  & \\
Data lake \& time-series databases & \checkmark & \checkmark &  & \\
Forecasting engine (sequence models)  &  &  & \checkmark & \checkmark \\
AI optimization algorithms   &  &  & \checkmark & \checkmark \\
Demand-response orchestration tools &   & \checkmark & \checkmark & \checkmark \\
External APIs (market, grid, weather) & \checkmark & \checkmark & \checkmark &  \\
User interfaces/dashboards (mobile apps, web)    &    &  &  & \checkmark \\
Automated control modules &  &    & \checkmark & \checkmark \\
\bottomrule
\end{tabular}
\end{table*}
This systematic mapping confirms that the ODP's technology stack, with its integrated AI engine, comprehensively addresses the functional requirements for data acquisition, system integration, intelligent analysis and control, and user interaction. This forms a robust foundation for optimizing energy and mobility services, creating synergistic benefits by enabling real-time, coordinated actions across the energy and transport domains that are not possible with siloed systems.

\subsection*{Step 3: Map functions to benefits}\label{meth_step3}
This step identifies the specific categories of benefits generated by each core ODP function. The anticipated benefits are categorized into three primary domains:\\
\textit{i) Economic benefits}: Direct financial gains or cost savings, encompassing reduced operational expenditures for grid operators and end-users, new revenue streams from energy services, enhanced energy efficiency, and optimized asset utilization.\\
\textit{ii) Reliability benefits}: Improvements in power system reliability and resilience, leading to reduced outage costs for consumers and businesses, enhanced service quality for end-users, and potentially deferred or optimized investments in grid infrastructure due to better load management.\\
\textit{iii) Environmental benefits}: Reductions in greenhouse gas (GHG) emissions (primarily CO$_2$) and local air pollutants (SOx, NOx, PM$_{2.5}$), contributing to climate change mitigation, improved public health, and greater integration of clean renewable energy.\\
Each ODP function (A-D) contributes to one or more of these benefit categories. For example, advanced data collection and processing (A) provides the necessary real-time data for precise grid state estimation and load forecasting, enabling more efficient asset utilization and informed decision-making, which directly translates to economic efficiency and enhanced system reliability. Enhanced interoperability (B) enables dynamic coordination between different market participants and grid entities, enabling new revenue streams from flexible assets and reducing system imbalances. The core AI-driven forecasting, optimization, and control functions (C) are pivotal in generating significant economic benefits (e.g., reduced energy procurement costs through peak load shaving and optimized energy trading facilitated by highly accurate price predictions, and increased revenues from ancillary service provision). These AI functions also yield substantial reliability benefits (e.g., prevention of grid overloads, improved voltage/frequency stability through predictive control, and reduced RES curtailment by matching demand with variable generation). Furthermore, these AI functions directly contribute to environmental benefits by maximizing the use of RES, optimizing EV charging for lower emission periods, and reducing fossil fuel-based generation for balancing. Decision support and automation (D) translates AI-generated insights into tangible outcomes, yielding direct economic benefits for users (e.g., lower charging costs due to optimized schedules, time savings from efficient routing) and operators (e.g., automated response to grid events), and indirectly fostering environmental benefits by promoting behaviors aligned with RES availability and lower carbon intensity.
Table \ref{tab_func-benefit} qualitatively summarizes how the ODP’s key functions map to specific benefit mechanisms.

\begin{table}[htbp]
\centering
\caption{Contributions of ODP functions to benefit categories (\cmark indicates a primary contribution, (i) indicates an indirect contribution).}
\label{tab_func-benefit}
\begin{tabular}{@{}lccc@{}}
\toprule
\textbf{Function} & \textbf{Economic} & \textbf{Reliability} & \textbf{Environmental} \\
\midrule
A. Advanced data \& info. collection/processing & \cmark & \cmark & (i) \\
B. Enhanced interoperability/system integration & \cmark & \cmark & \cmark \\
C. AI forecasting, optimization \& control & \cmark & \cmark & \cmark \\
D. Decision support \& automation & \cmark & (i) & \cmark \\
\bottomrule
\end{tabular}
\end{table}

This mapping highlights that economic benefits are pervasively driven by all functional areas, with particularly strong and direct contributions from the AI engine (Function C) and effective user/system engagement through transparent decision support (Function D). Reliability benefits primarily stem from improved situational awareness (Function A), seamless integration enabling coordinated responses (Function B), and intelligent grid management capabilities (Function C) that anticipate and mitigate grid issues. Environmental benefits accrue mainly from functions that facilitate higher RES penetration and smarter, cleaner energy consumption patterns for EVs/ETs (Functions B, C, and D), minimizing reliance on fossil fuels. 

\subsection*{Step 4: Monetize benefits}\label{meth_step4}
As described in Step 1, all benefits of the ODP-enhanced system are measured as the difference between the system with ODP and a counterfactual baseline (without ODP). This is done by defining a physical delta relative to the baseline and applying a valuation.
\[\Delta B_{k,t} = \left[\mathcal{M}_k(\text{ODP},t)-\mathcal{M}_k(\text{Baseline},t)\right] P_{k,t},\]
where $P_{k,t}$ is a monetary valuation of the delta-benefit, such as price.

The CBA adopts a 10‐year appraisal window (2026–2035) appropriate for an AI‐centric ODP whose enabling technologies undergo major upgrades within $\sim$5–10 years. This aligns with prior digital/smart-grid CBAs \citep{vilaplana2025dynamic}. Comparable platform-evaluation work reaches the same practical conclusion \citep{ruutu2017development}, and the shorter horizon avoids speculative long-term assumptions typical of structural assets. The choice of a 10-year horizon is therefore conservative: some shared digital-infrastructure costs are front-loaded, while several integration benefits continue to accrue beyond the first platform-refresh cycle, so a longer horizon would be expected to strengthen rather than weaken the investment case. To avoid reporting undocumented extensions, however, the quantitative results presented here are confined to the 2026–2035 window. Specific metrics, mathematical formulations, and assumptions are established for each benefit category. The details are presented in \ref{app_formulas}, based upon standard valuation methodologies \citep{ECJRC2012}, relevant market evidence \citep{BNEF_V2G_Report_2022}, and country-specific data. For each relevant benefit, the improvement margin over the baseline is directly tied to the enhanced capabilities, either superior forecasting accuracy or more efficient operational optimization, provided by the specific AI components of the ODP. 
To demonstrate the ODP's impact in varied national contexts, the proposed CBA framework is applied to Austria (AT), Hungary (HU), and Slovenia (SI). These countries offer a compelling mix of varying grid structures, renewable energy penetration levels, and electric mobility adoption trajectories, providing a heterogeneous Central European testbed for the ODP's cross-border and multi-sectoral capabilities.
\begin{table}[htbp]
\centering
\caption{Key country-specific assumptions for CBA (Illustrative summary for 2026 baseline and growth).}
\label{tab_country_assumptions_revised}
\resizebox{\columnwidth}{!}{
\begin{tabular}{@{}lcccP{0.3\textwidth}@{}}
\toprule
\textbf{Parameter} & \textbf{Austria (AT)} & \textbf{Hungary (HU)} & \textbf{Slovenia (SI)} & \textbf{Source/Justification} \\
\midrule
EV stock (2026, thousands) & 343.8 & 100.8 & 50.0 & National Registries, \citep{StatistaEV2023} \\
Annual EV Growth rate (CAGR 2026-35) & \SI{16.0}{\percent} & \SI{17.9}{\percent} & \SI{15.5}{\percent} & NECPs, \citep{BloombergNEF_EVOutlook2023} \\
ET stock (2026, thousands) & 4.4 & 0.4 & 0.3 & Pilot data, \citep{ACEA_TruckData2023} \\
Annual ET growth rate (CAGR 2026-35) & \SI{13.0}{\percent} & \SI{13.0}{\percent} & \SI{10.5}{\percent} & Industry projections, \citep{McKinseyElectricTrucks2022} \\
RES capacity (Solar + Wind, 2026 GW) & 10.5 & 8.0 & 1.5 & National TSO data, \citep{ENTSOE_TYNDP2022} \\
Annual RES addition (MW/yr avg.) & 664 & 2750 & 150 & NECP targets, \citep{SolarPowerEuropeOutlook2023} \\
Avg. electricity retail price (EV, €/kWh) & 0.22 & 0.15 & 0.18 & \citep{EurostatEnergyPrices2023}, adj. for EV tariffs \\
Grid CO$_2$ intensity (2026, g/kWh) & 120 & 280 & 190 & \citep{EEA_GHG_Intensity2023}, projected \\
Value of lost load (VoLL, €/kWh) & 10.0 & 7.5 & 8.0 & Nat. regulator studies, \citep{ACER_VoLL_Report2021} \\
Battery life Ext. (smart charging) & \multicolumn{3}{c}{\SIrange{8}{12}{\percent} reduction in degradation rate} & \citep{maheshwari2020optimizing} \\
Social cost of carbon (€/tCO$_2$) & \multicolumn{3}{c}{Starts at €85 in 2026, rising to €120 by 2035} & EU guidance \citep{EUCarbonValuation2022}, linear interpolation. \\
\bottomrule
\end{tabular}%
}
\end{table}
Table \ref{tab_country_assumptions_revised} summarizes key country-specific input assumptions for the baseline year 2026 and projected growth rates for the analysis period (details are presented in \ref{app_annual_proj}). These assumptions are derived from recent national energy and climate plans (NECPs), statistical office data \citep{EurostatData2023}, reports from international agencies like the IEA \citep{IEA_CountryReports_2023}, and specialized consultancies (e.g., BloombergNEF, McKinsey) to ensure their credibility and relevance to the respective national contexts. For instance, the EV/ET growth rates reflect ambitious but achievable national targets and industry projections, while RES capacity additions align with NECP goals.
Battery-health benefits are monetized as the NPV of the modeled reduction in degradation cost relative to the without-ODP baseline. Empirical studies report 8--12\% lower degradation under health-aware charging protocols: narrower SoC windows, C-rate limits, and temperature-aware scheduling \citep{maheshwari2020optimizing}. This paper uses the lower bound of that range (8\%) as a conservative benchmark.
The benefit model is operationalized through eight streams with explicit physical drivers, valuation factors, and equation-level traceability ( \ref{app_formulas_rod}--\ref{app_formulas_rap}). For illustration, the avoided energy curtailment (AEC) benefit---the largest single stream---is computed as
\begin{equation}
\mathrm{AEC}_{c,t} = \bigl( \mathrm{RES}^{\mathrm{ODP}}_{\mathrm{gen},c,t} 
      - \mathrm{RES}^{\mathrm{baseline}}_{\mathrm{gen},c,t} \bigr) 
    \times P_{\mathrm{wholesale},c,t} \times (1 - \text{loss factor}),
\end{equation}
where the ODP delta arises from improved 15-min to 72-h RES forecasting and real-time charging  alignment (literature benchmark: 10--15\,\% curtailment reduction). Full equations for all eight streams, valuation parameters, and accounting-order logic are provided in \ref{app_formulas}.

To prevent double counting, each marginal MWh of flexibility is assigned to a single primary monetization channel in each time step using a fixed accounting order: (1) reliability and congestion relief (ROD/GSMS), (2) RES absorption that would otherwise be curtailed (AEC), (3) market arbitrage and ancillary services (ROETAS), and (4) residual demand-response and peak-load benefits (CSDR-PLR). Any value already captured in an upstream channel is removed from downstream calculations. FES is restricted to vehicle-level operating and battery-degradation savings not already represented in system-level streams. 
 \begin{itemize}
\item \textbf{Reduction in operational downtime (ROD).} ROD captures avoided unserved energy and service interruptions due to better fault anticipation and faster restoration (Eq.~\ref{eq_rod_appendix}). The physical driver is reduced outage exposure under AI-enabled predictive maintenance and situational awareness; valuation uses avoided outage-cost factors.
\item \textbf{Revenue from optimized trading and ancillary services (ROETAS).} ROETAS captures incremental market value from improved price/imbalance forecasting and optimized flexibility bids. (Eq.~\ref{eq_roetas_appendix}). The stream includes both arbitrage margins and net ancillary-service revenues relative to baseline operations. 
\item \textbf{Cost savings from demand response and peak-load reduction (CSDR-PLR).} CSDR-PLR quantifies avoided system costs from coordinated charging and load shifting (Eq.~\ref{eq_csdrplr_appendix}). The physical mechanism is peak flattening and reduced balancing stress under ODP dispatch logic
 \item \textbf{Avoided energy curtailment (AEC).} AEC monetizes renewable output that is utilized instead of curtailed through forecast-informed flexibility activation (Eq.~\ref{eq_aec_appendix}). This stream links RES prediction quality with controllable-demand alignment.
 \item \textbf{Fleet energy and operational savings (FES).} FES captures EV/ET operating savings from route-energy co-optimization and health-aware charging (Eq.~\ref{eq_fes_appendix}). It includes electricity-consumption savings and battery-degradation cost reductions under constrained charging policies.
\item \textbf{Grid stability and management savings (GSMS).} GSMS monetizes avoided DSO/TSO operational expenditures related to congestion mitigation, balancing actions, and reliability management (Eq.~\ref{eq_gsms_appendix}). The physical basis is reduced expected stress events under improved forecast-control coordination. 
\item \textbf{CO$_2$ reduction benefits.} This stream monetizes net greenhouse-gas reductions associated with higher RES utilization, better load timing, and lower balancing emissions (Eq.~\ref{eq_co2_appendix_refined}). The magnitude of this stream depends critically on the marginal CO$_2$ emission factor (MEF) at the hours when ODP shifts charging load. 
\item \textbf{Reduction in air pollutants (RAP).} RAP monetizes health and environmental gains from lower NO$_x$, SO$_x$, and PM$_{2.5}$ emissions (Eq.~\ref{eq_rap_appendix}), applying pollutant-specific valuation factors.
 \end{itemize}
 
\begin{figure}[htbp]
\centering
\includegraphics[width=0.75\linewidth]{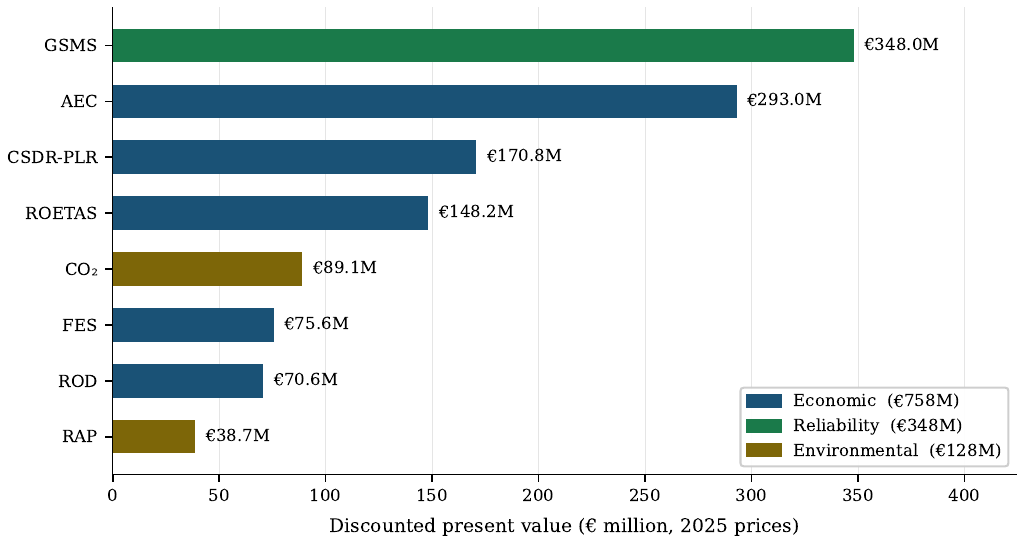}
\includegraphics[width=0.75\linewidth]{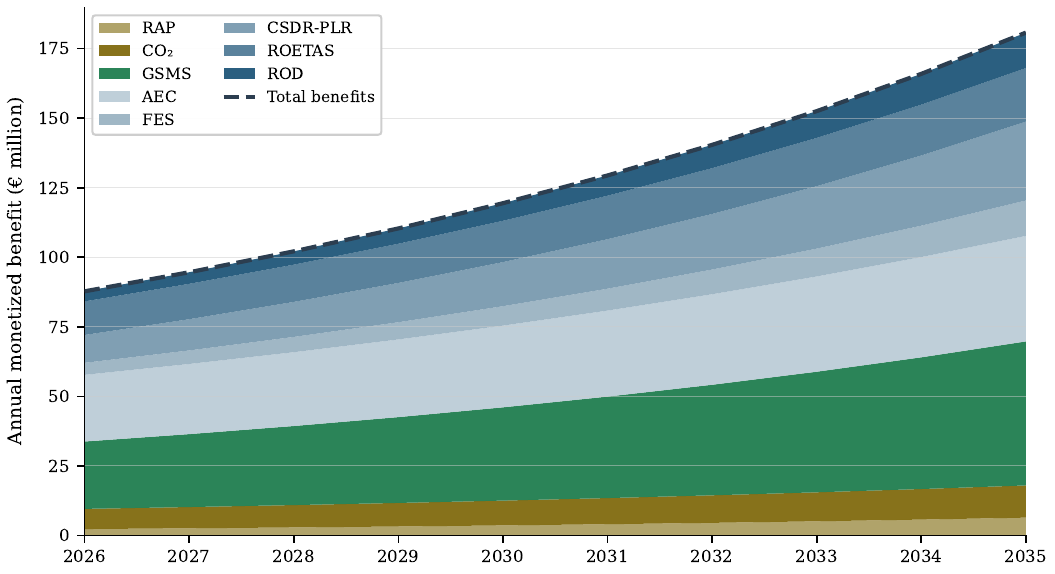}\\
\includegraphics[width=0.85\linewidth]{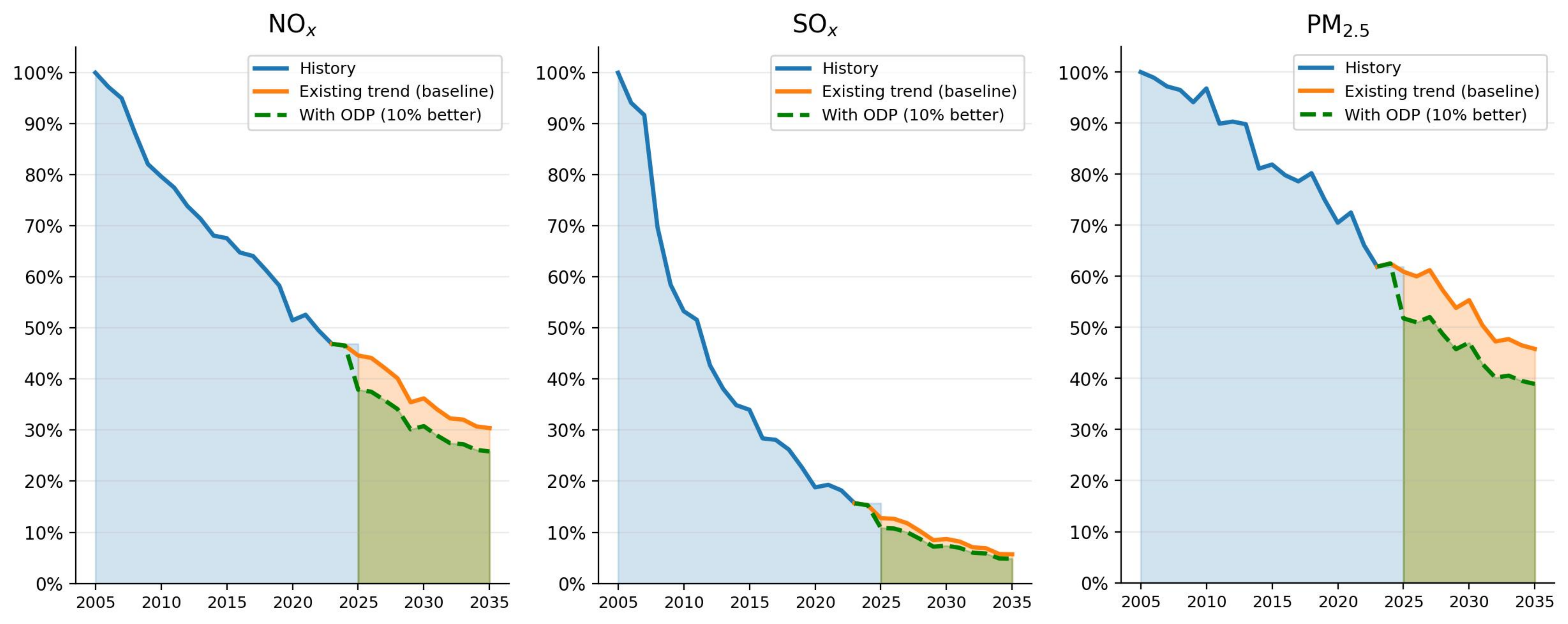}
\caption{Benefit analysis overview. \textbf{Top}: total discounted present value of each benefit stream over 2026--2035, aggregated for AT, HU, and SI (colour-coded by category: blue = economic, green = reliability, amber = environmental). \textbf{Middle}: year-on-year accumulation of all eight benefit streams, showing compound annual growth of approximately 7.9\% as EV/ET penetration and RES capacity expand. \textbf{Bottom panel}: NO$_x$, SO$_x$, and PM$_{2.5}$ emission trajectories for the baseline and ODP-enhanced scenarios, indexed to EU-27 proxy (2005 = 100\%), demonstrating the platform's contribution to meeting EU air-quality objectives alongside decarbonization targets. All values discounted at 4 \% social rate to 2025; see \ref{app_details} for exact formulas.}
\label{fig_benefit-breakdown_revised}
\end{figure}

Across all streams, annual benefits are computed as baseline-to-ODP deltas, then discounted and aggregated across countries. This ensures methodological consistency between architecture-level mechanisms and economic outputs.
Table \ref{tab_benefits-summary_revised} presents the total present value of benefits over $2026-2035$, aggregated across the three countries and categorized by type. Detailed annual benefit projections, underpinning these present values, are also provided in \ref{app_annual_proj}.
\begin{table*}[htbp]
\centering
\caption{Aggregated projected benefits of the ODP across Austria, Hungary, and Slovenia over 10 years (2026-2035, present values in € million, all values discounted at 4 \% social rate to 2025; see \ref{app_details} for exact formulas.).}
\label{tab_benefits-summary_revised}
\resizebox{\textwidth}{!}{
  \begin{threeparttable}
  \begin{tabular}{@{}P{0.35\textwidth} r r S[table-format=1.1]@{}}
  \toprule
  \textbf{Benefit category} & \textbf{Component} & \textbf{Present value (€ million)} & \textbf{Share of total benefits (\%)} \\
  \midrule
  \multirow{5}{=}{Economic benefits}
   & Reduction in operational downtime (ROD) & 70.63 & 5.7 \\
   & Optimized energy trading \& ancillary services (ROETAS) & 148.20 & 12.0 \\
   & Demand response \& peak load reduction (CSDR-PLR) & 170.80 & 13.8 \\
   & Fleet energy \& operational savings (FES) & 75.61 & 6.1 \\
   & Avoided energy curtailment (AEC) & 292.99 & 23.7 \\
  \cmidrule{2-4}
   & \textit{Subtotal economic benefits} & \textit{758.23} & \textit{61.5} \\
  \midrule
  \multirow{1}{=}{Reliability benefits\tnote{a}}
   & Grid stability \& management savings (GSMS)\tnote{b} & 347.99 & 28.2 \\
  \cmidrule{2-4}
   & \textit{Subtotal reliability benefits} & \textit{347.99} & \textit{28.2} \\
  \midrule
  \multirow{2}{=}{Environmental benefits}
   & CO$_2$ emission reduction benefits & 89.07 & 7.2 \\
   & Reduction in air pollutant benefits (RAP) & 38.64 & 3.1 \\
  \cmidrule{2-4}
   & \textit{Subtotal environmental benefits} & \textit{127.71} & \textit{10.4} \\
  \midrule
  \textbf{Total monetized benefits (PV)} & & \textbf{1233.9} & \textbf{100.0} \\
  \bottomrule
  \end{tabular}
  \begin{tablenotes}
  \footnotesize
  \item[a] Reliability benefits are substantial. GSMS captures direct grid operational savings. ROD, while categorized under economic, also reflects improved reliability. AEC enhances stability by better integrating variable RES.
  \item[b] GSMS: Reduced costs for grid regulation, balancing services (not otherwise captured in ROETAS), and congestion management due to ODP's predictive control and enhanced visibility.
  \end{tablenotes}
  \end{threeparttable}}
\end{table*}
Economic benefits constitute the largest share (61.5\% of total benefits). Within this category, avoided energy curtailment (AEC) is the single largest stream at \euro{}293M (23.7\%), reflecting the ODP's ability to align flexible EV/ET charging with periods of high RES surplus. Grid stability and management savings (GSMS, \euro{}348M) form the dominant reliability stream, underscoring the platform's role as a non-wires alternative that defers or avoids conventional congestion-management expenditures. Fig.~\ref{fig_benefit-breakdown_revised} shows both the aggregate breakdown and the year-on-year growth trajectory: benefits compound as fleet penetration and RES capacity expand over the horizon, with total annual benefits growing from \euro{}81M in 2026 to \euro{}174M in 2035.  Environmental benefits (10.4\%, \euro{}128M) stem from CO$_2$ reductions (\euro{}89M) and local air-pollutant abatement (\euro{}39M), both amplified by the ODP's ability to dispatch EV/ET charging into low-carbon hours. Fig.~\ref{diurnal_profile} illustrates this mechanism: the platform consistently shifts charging load away from high marginal-emission-factor (MEF) periods, particularly in Hungary where grid carbon intensity exceeds 0.28\,tCO$_2$/MWh during peak fossil dispatch hours.

This enhancement increases the \textit{granularity and policy relevance} of the environmental assessment without altering the fundamental framework for benefit calculation. Fig. \ref{fig_benefit-breakdown_revised} presents total benefits by category and their year-by-year evolution. The top-left panel displays the total economic, reliability, and environmental benefits; the top-right panel shows the annual trajectory of each benefit over the project horizon. The bottom panel plots the air-pollutant trends under the existing system and the ODP-enhanced system.

The stacked-area chart (Fig.~\ref{fig_benefit-breakdown_revised}, right) depicts the year-on-year accumulation of all eight benefit streams. AEC and GSMS dominate throughout the horizon, together accounting for over 52\% of annual benefits in every year. The remaining streams grow steadily as EV/ET adoption expands, with FES and CSDR-PLR exhibiting the steepest trajectories because they scale directly with fleet size and charging capacity (Fig.~\ref{projections}). At the country level, benefit composition differs systematically. Austria's benefit profile is led by FES and ROD, reflecting its larger and more mature EV/ET fleet. Hungary's profile is dominated by AEC, driven by the rapid solar expansion that creates frequent curtailment risk. Slovenia's relatively small scale is offset by strong grid-flexibility and ET-logistics gains, yielding a BCR of 1.41 that matches the aggregated result. These country-specific patterns are discussed further in Section~\ref{dis} and summarized quantitatively in Table~\ref{tab_app_country_mcs_populated} and Fig.~\ref{comparison}.

\begin{figure}[htbp]
\centering
\includegraphics[width=0.95\linewidth]{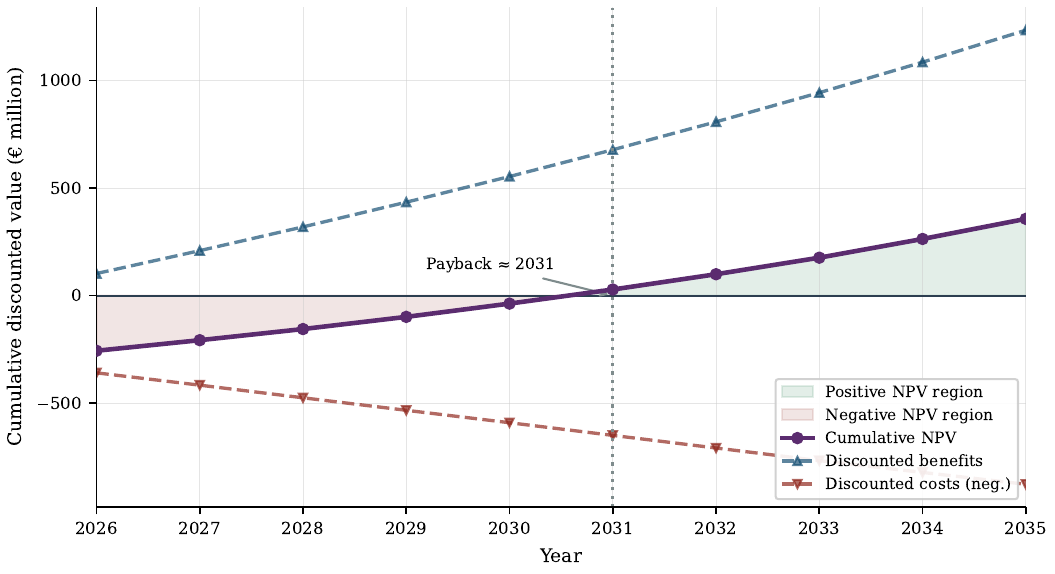}
\caption{Cumulative discounted NPV trajectory and payback analysis (AT, HU, SI combined). Benefits are discounted at 4\% and accumulated annually; the intersection with the x-axis identifies the approximate payback year. The growing positive surplus from 2031 onward confirms that the platform generates compounding net value over the analysis horizon.}
\label{benefit_payback}
\end{figure}

Fig.~\ref{benefit_payback} presents the cumulative discounted NPV trajectory. Benefits are discounted at 4\% and accumulated year by year; the project crosses into positive cumulative NPV territory in 2031, confirming an approximately six-year payback period from the start of operations. The growing positive surplus after 2031 demonstrates that the platform generates compounding net value in the later years of the analysis horizon, driven by rising benefit streams against a stabilizing cost base.

\subsection*{Step 5: Quantify costs}\label{meth_step5}
\begin{figure}[htbp]
\centering
\includegraphics[width=0.9\linewidth]{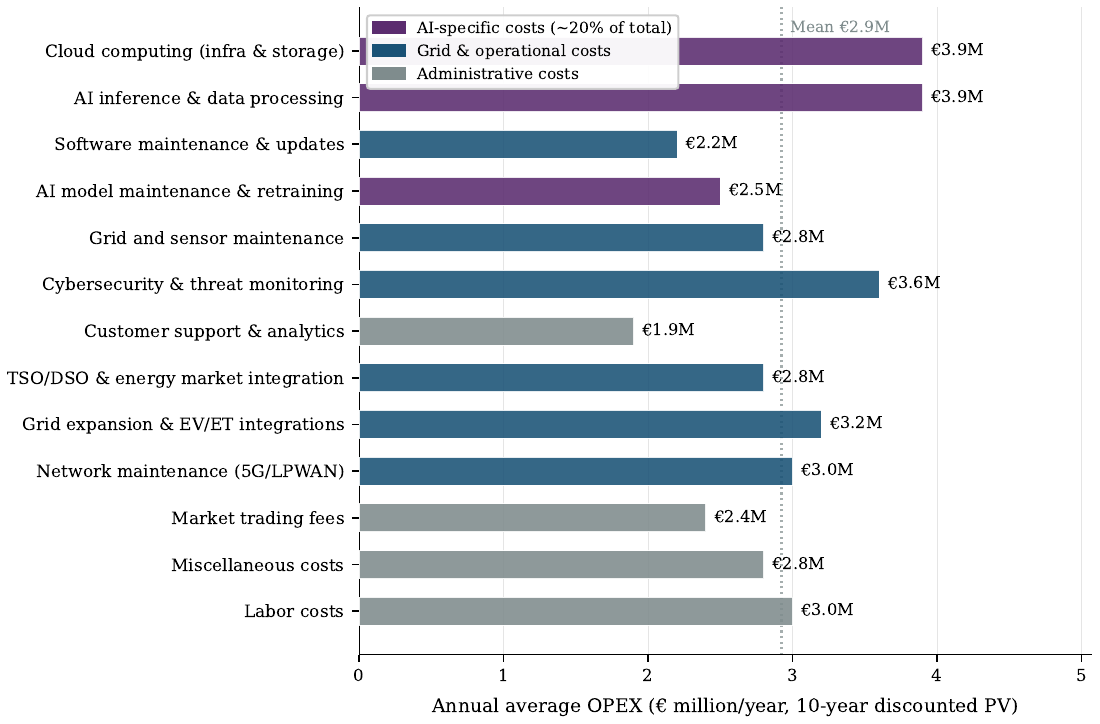}
\caption{Indicative OPEX distribution by cost category in the first year of operation (2026), aggregated for AT, HU, SI. Major categories include cloud computing (€3.9M), AI inference \& data processing (€3.9M), cybersecurity and threat monitoring (€3.6M), labor costs (€3.0M), grid \& sensor maintenance (€2.8M), and AI model maintenance \& retraining (€2.5M). Total discounted 10-year OPEX is €315.9 million.}
\label{fig_opex_revised}
\end{figure}

This step provides a comprehensive assessment and quantification of all costs associated with implementing and operating the ODP over 2026–2035. Costs are categorized into capital expenditures (CAPEX) and operational expenditures (OPEX). Detailed breakdowns that differentiate initial platform development from ongoing integration and operation are provided in \ref{app_capex} (CAPEX) and \ref{app_opex} (OPEX). Below, we summarize their main elements:

 \textbf{CAPEX}: These are primarily upfront investments incurred during the initial deployment phase (2026) and for system expansion in subsequent years. Major CAPEX components include hardware procurement (sensors, servers, communication devices), core software development for the ODP platform, and crucially, the initial development and training of the AI engine's specialized models. System integration and project management costs are also included. A significant portion of CAPEX is incremental, related to the integration of new EVs, ETs, and RES installations annually into the ODP-managed ecosystem. Our estimates indicate a total discounted CAPEX of €561.3 million over the 10-year period, with approximately 15\% of the initial platform CAPEX directly attributable to AI model development and associated infrastructure (e.g., specialized GPU clusters for training).\\
\textbf{OPEX}: These are recurring annual costs necessary to operate and maintain the ODP. Key OPEX categories include cloud computing resources for AI inference and data processing, ongoing software maintenance (including continuous AI model monitoring, re-training pipelines to counter data/concept drift, and performance optimization), communication fees, robust cybersecurity operations, labor costs for technical staff (e.g., data scientists, AI engineers, system administrators), and hardware maintenance. Annual OPEX is projected to start at approximately €31.6 million (discounted) in the first year (2026), with an estimated 20\% of this directly related to AI operations and maintenance. The total discounted OPEX over 10 years is estimated at €315.9 million.\\
Fig.~\ref{fig_opex_revised} shows the first-year (2026) OPEX breakdown across 13 categories. Cloud computing and AI inference together account for \euro{}7.8M --- the two largest individual line items --- confirming that sustaining the AI engine's real-time inference capability is the primary recurring cost driver. Cybersecurity operations (\euro{}3.6M) rank third, reflecting the critical-infrastructure status of an integrated energy-mobility platform. Labor costs (\euro{}3.0M) cover the data scientists, AI engineers, and system administrators required for continuous model monitoring and retraining. Across all 13 categories, AI-specific items (cloud, inference, model maintenance) represent approximately 30\% of first-year OPEX, declining modestly as a share in later years as operational processes mature. 

The total discounted investment cost (CAPEX + OPEX) for the ODP deployment and operation over the decade is approximately €877.2 million. Cost estimates are benchmarked against similar smart grid and digital platform deployments in the energy sector \citep{Stromsather2018, SmartEnergyPlatformsReport2022} and include reasonable contingencies for unforeseen expenditures. The specific breakdown for AI-related costs captures both initial development (CAPEX) and ongoing operational and maintenance expenses (OPEX), which are essential for sustaining the performance of AI models in a dynamic environment.

CAPEX for shared platform development and core infrastructure, including AI engine development, was allocated to each country based on a composite weighting factor considering projected deployment scale (e.g., number of managed EVs/ETs, RES capacity integrated, and overall energy consumption), as detailed in \ref{app_capex} and \ref{app_opex}. As a result, Austria accounts for approximately 45\% of total costs, Hungary 35\%, and Slovenia 20\%. Detailed cost breakdowns, including specific required items for AI system development and ongoing AI model maintenance, are comprehensively presented in \ref{app_capex} and \ref{app_opex}, with annual cost projections provided in \ref{app_annual_proj}.

\subsection*{Step 6: Compare costs and benefits}\label{meth_step6}
In this step, the total present value of monetized benefits is compared with the total present value of costs. Key financial performance indicators---notably NPV and BCR---are computed to evaluate the economic viability of ODP deployment under the baseline assumptions.
Table \ref{tab_comparison_revised} summarizes the aggregated financial outcomes over the 10-year analysis horizon, discounted at 4\%.

\begin{table}[htbp]\centering
\caption{Comparison of total benefits and costs (2026--2035, discounted at 4\% to 2025, aggregated for AT, HU, SI).}
\label{tab_comparison_revised}
\begin{tabular}{@{}l S[table-format=4.1]@{}}
\toprule
\textbf{Category} & \textbf{Present value (€ million)} \\
\midrule
Total benefits & 1233.9 \\
CAPEX & 561.3 \\
OPEX & 315.9 \\
Total costs (CAPEX + OPEX) & 877.2 \\
\midrule
\textbf{NPV} (Total benefits $-$ Total costs) & \textbf{356.7} \\
\textbf{BCR} (Total benefits / Total costs) & \textbf{1.41} \\
\bottomrule
\end{tabular}
\end{table}

\begin{figure}[t!]
\centering
\includegraphics[width=0.9\linewidth]{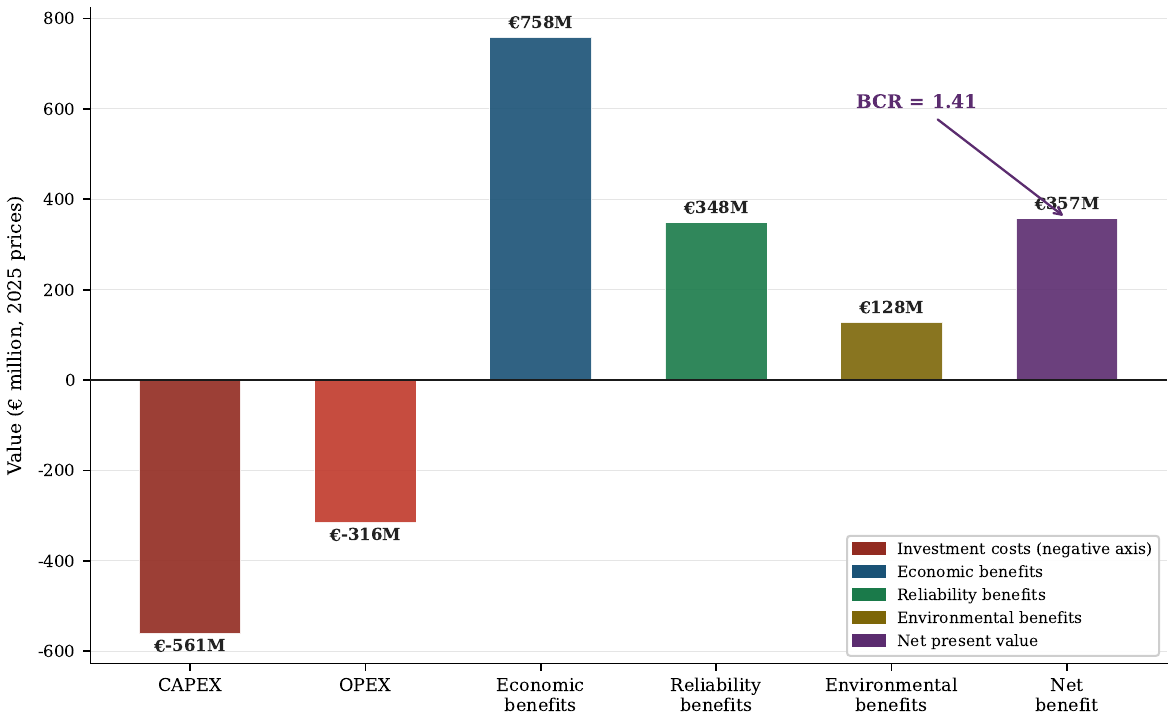}
\caption{Cost-benefit comparison over 2026--2035 (discounted values, aggregated for AT, HU, SI). Total costs (CAPEX and OPEX) are shown as negative bars. Total benefits are categorized into Economic, Reliability, and Environmental categories, each represented by distinct colors. The NPV is significantly positive.}
\label{fig_cost-benefit_revised}
\end{figure}

\begin{figure}[htbp]
\centering
\includegraphics[width=0.95\linewidth]{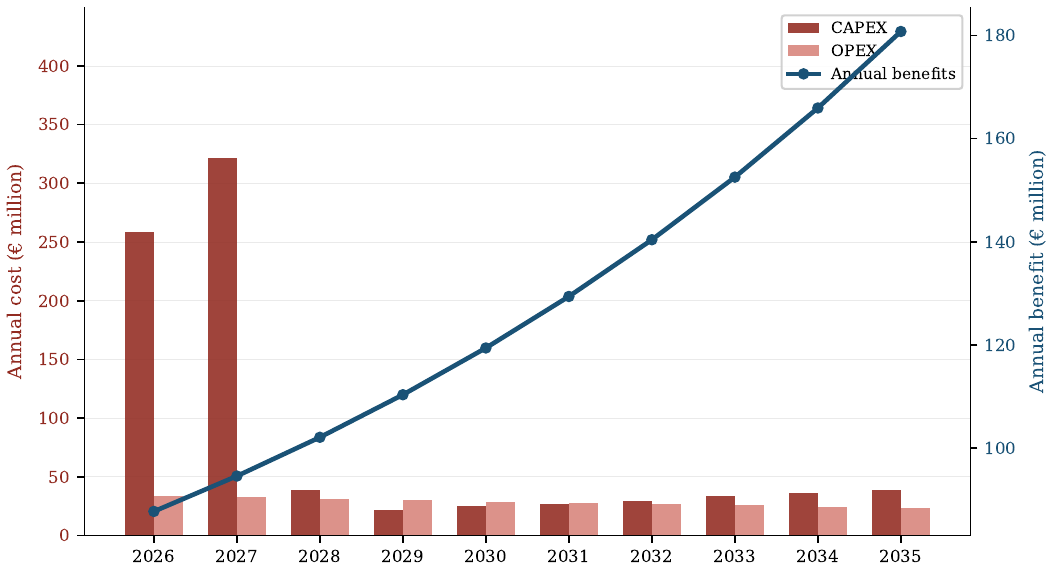}
\caption{Annual cost profile versus benefit trajectory, 2026--2035. Left axis (bars): CAPEX (dark red, front-loaded in 2026) and OPEX (lighter red, annual recurring). Right axis (line): total undiscounted annual benefits, growing from \euro{}81M in 2026 to \euro{}174M in 2035 as fleet penetration and RES capacity expand. The crossing of the benefit line above annual total costs around 2030 marks the annual cash-flow breakeven point.}
\label{annual_cost_benefit}
\end{figure}

Total discounted benefits (\euro{}1,233.9M) exceed total discounted costs (\euro{}877.2M) by \euro{}356.7M, yielding a BCR of 1.41. Every euro of investment generates \euro{}1.41 in discounted value, a result that is robust to substantial cost and benefit perturbations as shown in Step~7. Fig.~\ref{fig_cost-benefit_revised} illustrates this as a cost--benefit chart: CAPEX and OPEX bars extend below the zero line, while economic, reliability, and environmental benefit bars extend above it, landing on a net positive value. The payback curve in Fig.~\ref{benefit_payback} shows the project crossing into a positive cumulative NPV position in 2031, with the annual benefit trajectory (Fig.~\ref{annual_cost_benefit}) exceeding annual total costs from approximately 2030 onward. The annual benefit line grows steadily from \euro{}81M in 2026 to \euro{}174M in 2035 as fleet penetration and RES capacity compound, while annual OPEX stabilizes and CAPEX declines sharply after the 2026 deployment year. 
The positive NPV and BCR $>$ 1 strongly support the economic and financial feasibility of the ODP deployment. These results are based on conservative assumptions for benefit realization and cost estimations, detailed in \ref{app_capex} and \ref{app_opex}, and include the explicit quantification of AI's performance-driven contributions.

\subsection*{Step 7: Sensitivity analysis and risk assessment}\label{meth_step7}
To assess the robustness of the CBA results to inherent uncertainties in key input parameters, a comprehensive sensitivity and risk analysis was performed. This involved both scenario-based sensitivity testing and probabilistic risk analysis using Monte Carlo simulation.
First, scenario-based analyses were conducted by varying critical benefit and cost parameters. Table \ref{tab_scenarios_revised} presents the impact of these variations on the project's NPV and BCR. Crucially, this includes a specific scenario for AI adoption variability, where the effectiveness factors are reduced.

\begin{figure}[htbp]
\centering
\includegraphics[width=0.95\linewidth]{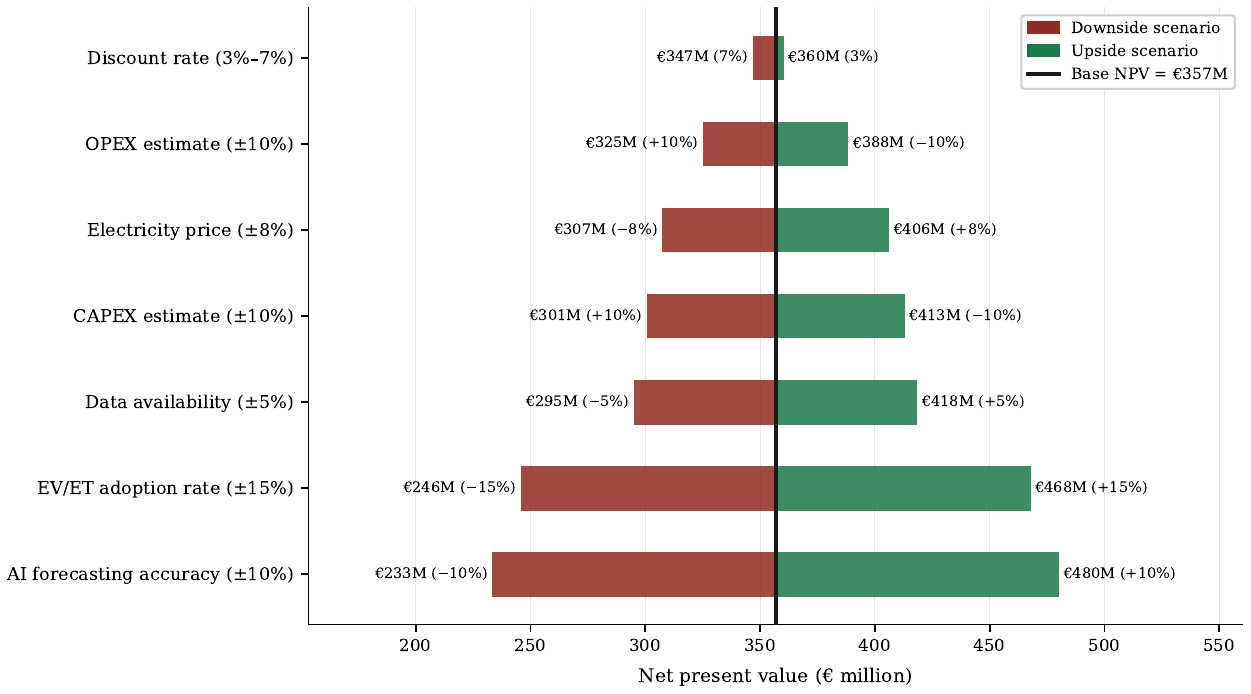}
\caption{Sensitivity analysis of NPV and BCR under uncertainty in CAPEX, OPEX, AI performance, data availability, EV adoption, and electricity prices.}
\label{tornado_sensitivity}
\end{figure}

\begin{table}[htbp]\centering
\caption{Sensitivity of NPV and BCR to variations in key benefit and cost parameters (aggregated for AT, HU, SI).}
\label{tab_scenarios_revised}
\begin{tabular}{@{}l S[table-format=3.1] S[table-format=1.2]@{}}
\toprule
\textbf{Scenario} & \textbf{NPV (€ million)} & \textbf{BCR} \\
\midrule
Base Case & 356.7 & 1.41 \\
Total benefits $+30\%$ & 480.2 & 1.55 \\
Total benefits $-30\%$ & 233.4 & 1.27 \\
Total costs $+20\%$ & 269.1 & 1.28 \\
Total costs $-20\%$ & 444.5 & 1.56 \\
Discount rate 3\% (instead of 4\%) & 360.3 & 1.42 \\
Discount rate 5\% (instead of 4\%) & 353.3 & 1.39 \\
\bottomrule
\end{tabular}
\end{table}
The tornado diagram (Fig.~\ref{tornado_sensitivity}) ranks parameters by their impact on NPV. AI forecasting accuracy and EV/ET adoption rate exert the widest influence, each capable of shifting NPV by over \euro{}120M in either direction under $\pm$10\% and $\pm$15\% perturbations respectively. CAPEX and OPEX perturbations have a narrower but still material effect (\euro{}56--112M range), while the discount rate sensitivity is comparatively modest (\euro{}13M over a 3--7\% range). Critically, even the most adverse combined scenario (reduced accuracy, lower adoption, higher costs) preserves a positive NPV, confirming a comfortable margin of safety. A Monte Carlo simulation (50,000 trials for the aggregated case) was then performed as a probabilistic risk-assessment framework. Country-level simulations reported later use 10,000 trials and are presented separately for comparability. Key parameters---EV/ET adoption rates, energy-price volatility, CAPEX/OPEX unit costs, AI model-performance factors, and data availability---are jointly sampled from calibrated distributions (normal for adoption and AI performance, triangular for price, and uniform for selected CAPEX/OPEX cost scalars).
\begin{figure}[htbp]
\centering
\includegraphics[width=0.95\linewidth]{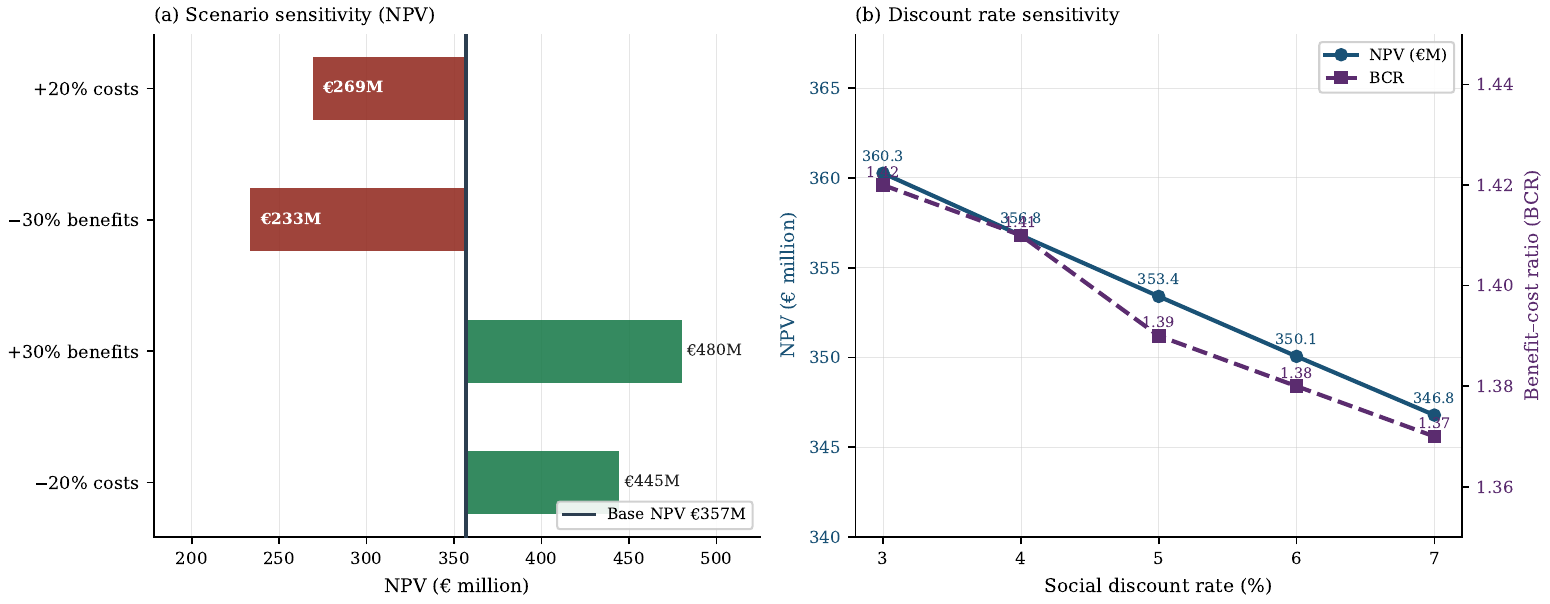}
\caption{Sensitivity analysis. Panel (a): scenario-based tornado chart showing NPV ranges when total benefits or total costs deviate by $\pm$20--30\% from the baseline (base NPV = \euro{}357M). All scenarios maintain a positive NPV, confirming a comfortable margin of safety. Panel (b): discount rate sensitivity showing that NPV declines from \euro{}360M at 3\% to \euro{}347M at 7\%, while the BCR remains above 1.37 throughout, demonstrating low sensitivity to the cost of capital.}
\label{sensitivity}
\end{figure}

\begin{figure}[htbp]
\centering
\includegraphics[width=1\linewidth]{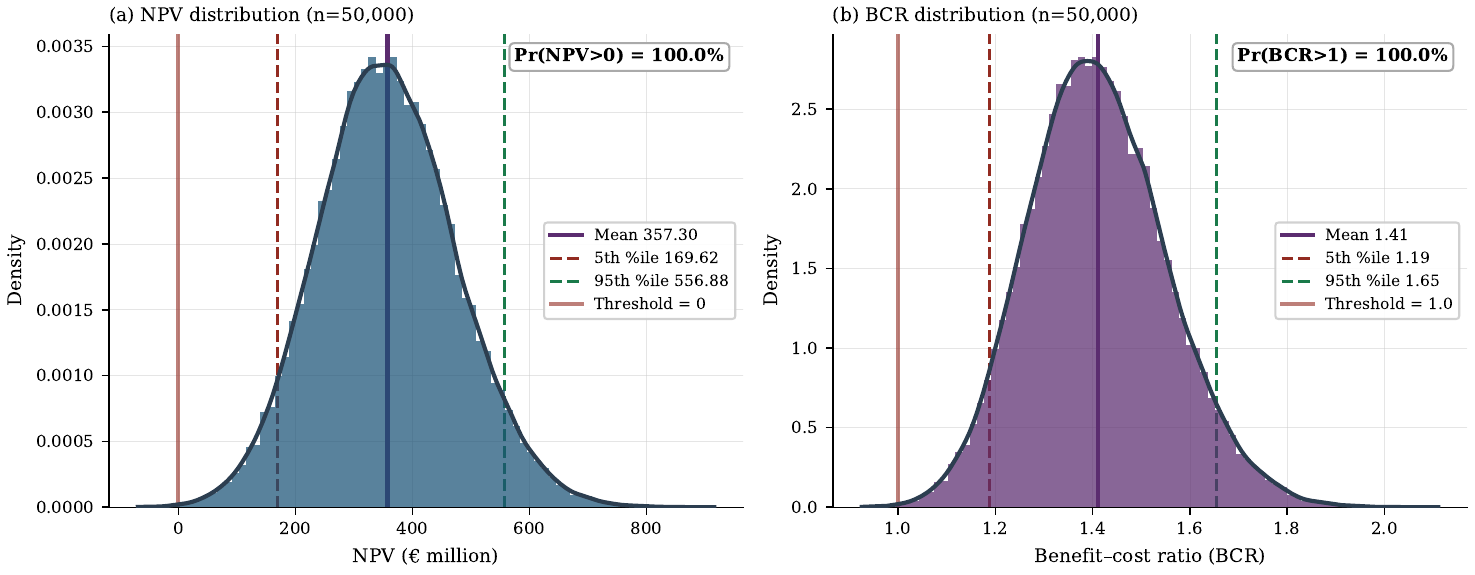}
\caption{Monte Carlo simulation results for aggregated NPV and BCR (50,000 trials for AT, HU, and SI combined). \textbf{Left:} Probability distribution of NPV (Million €); mean NPV is €357.30M, with simulated 5th and 95th percentiles of €169.62M and €556.88M. \textbf{Right:} Probability distribution of BCR; mean BCR is 1.41, with simulated 5th and 95th percentiles of 1.19 and 1.65. The figure also reports the share of trials satisfying NPV$>$0 and BCR$>$1.}
\label{fig_montecarlo_revised}
\end{figure}

Fig.~\ref{fig_montecarlo_revised} shows the probability distributions of NPV and BCR for the aggregated three-country case. The NPV distribution is concentrated in positive territory with a mean of \euro{}357.30M; the simulated 5th--95th percentile interval is [\euro{}169.62M, \euro{}556.88M], indicating substantial upside while retaining a positive lower-tail outcome. The BCR distribution is centred at 1.41 with a simulated 5th--95th percentile interval of [1.19, 1.65]. The simulations jointly perturb energy-price volatility, EV/ET adoption rates, AI forecasting accuracy, data availability, and CAPEX/OPEX around the baseline NPV and BCR estimates.

The Monte Carlo simulation yields a mean NPV of €357.30 million and a mean BCR of 1.41. Across the 50,000 aggregated trials, all simulated outcomes satisfy NPV$>$0 and BCR$>$1 for the sampled parameter distributions used here. The slight difference between the Monte Carlo mean NPV/BCR and the deterministic base-case values can arise from asymmetric probability distributions of input variables, non-linear relationships between inputs and output metrics, and correlations among the sampled variables, all of which are captured in the probabilistic analysis. These results indicate that investment viability is robust to plausible adverse combinations of adoption, cost, and performance shocks; however, uncertainty remains most sensitive to CAPEX realization and AI-performance degradation, which should be prioritized in implementation governance.

Fig.~\ref{tornado_sensitivity} (bottom panel) presents the extended one-way sensitivity analysis covering seven key parameters sorted by impact range. AI forecasting accuracy (±10\%) and EV/ET adoption rate (±15\%) generate the widest NPV ranges (\euro{}246M and \euro{}222M respectively), confirming that AI performance and fleet uptake are the primary value drivers. CAPEX variability (±10\%) produces a \euro{}112M range, while OPEX variation (±10\%) contributes a narrower \euro{}63M range. Discount rate sensitivity (3\%--7\%) is comparatively modest at \euro{}13M, indicating that the investment case is robust to the choice of social discount rate across the range of rates applied in EU project appraisals. Critically, even the most adverse single-parameter scenario (AI accuracy reduced by 10\%) preserves a positive NPV above \euro{}233M, confirming a comfortable margin of safety throughout the sensitivity space.  

\begin{figure}[htbp]
\centering
\includegraphics[width=0.95\linewidth]{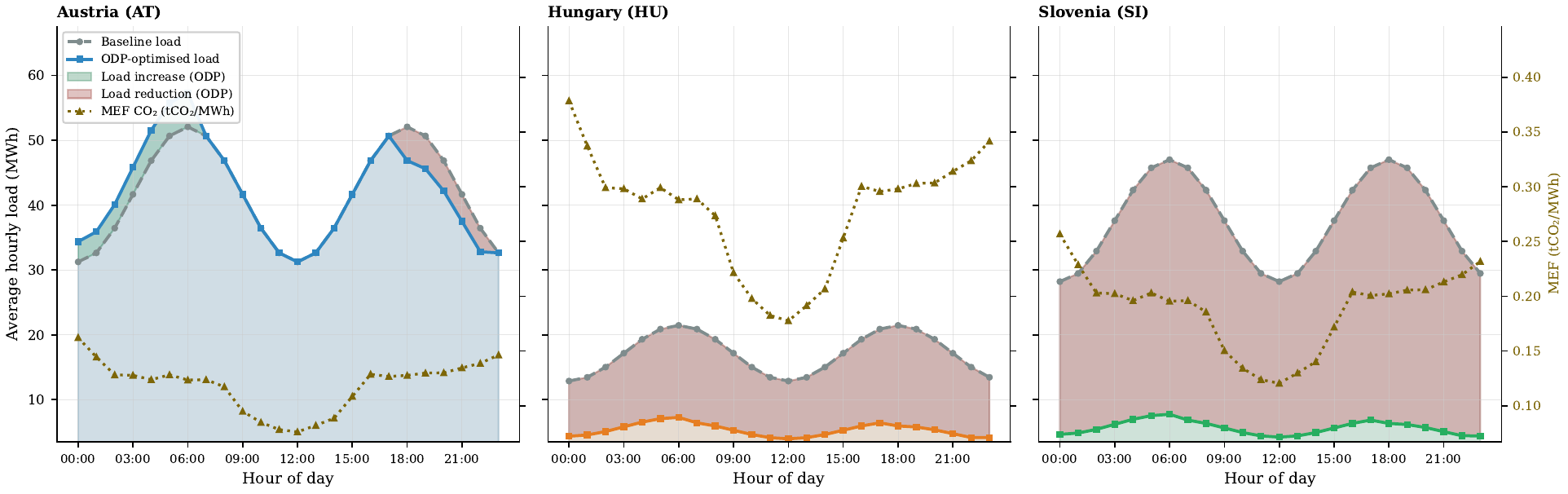}
\caption{Diurnal load profile (annual average, 2026) comparing baseline and ODP-optimised charging schedules, overlaid with the marginal CO$_2$ emission factor (MEF) for Austria, Hungary, and Slovenia. Green shading indicates hours where ODP shifts load upward to absorb cheap, low-carbon generation; red shading marks hours where ODP suppresses peak demand. Hungary's higher MEF curve reflects its greater coal dependence, while Slovenia's lower MEF reflects large hydro share. ODP consistently shifts charging into low-MEF hours, amplifying CO$_2$ co-benefits.}
\label{diurnal_profile}
\end{figure}

\begin{figure}[htbp]
\centering
\includegraphics[width=0.95\linewidth]{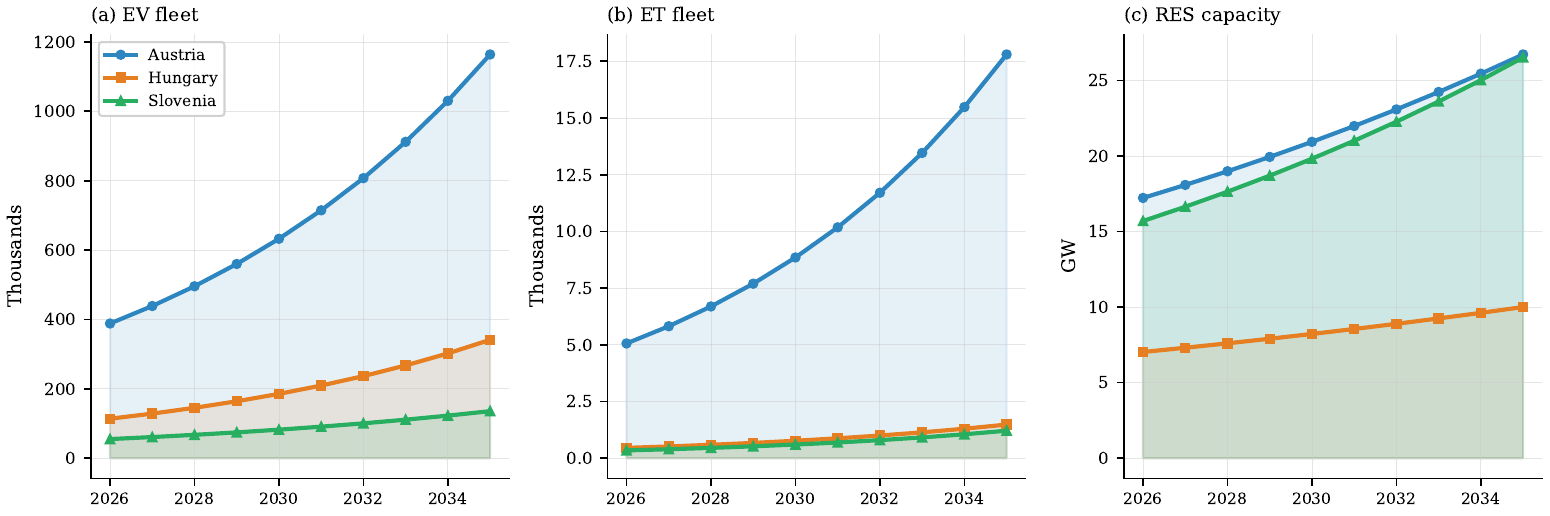}
\caption{Annual projections of EV fleet (thousands), ET fleet (thousands), and RES capacity (GW) for Austria, Hungary, and Slovenia, 2026--2035. Austria maintains the largest EV and ET fleets throughout the horizon, while Hungary exhibits the fastest RES-capacity expansion. Slovenia grows from a smaller base across all three dimensions. Values are consistent with Tables~B.15 and~B.16.}
\label{projections}
\end{figure}

Qualitative risk assessment identified additional factors: regulatory and policy risks (e.g., changes in market design or data sharing regulations), cybersecurity and data privacy risks (given the sensitive nature of energy and mobility data), technology adoption and integration risks (e.g., slower than expected stakeholder uptake, challenges in integrating with legacy systems), and stakeholder acceptance. Mitigation strategies include proactive engagement with policymakers to shape supportive regulatory frameworks, robust cybersecurity measures that adhere to international standards such as ISO 27001 and GDPR compliance, a phased deployment approach to manage technical complexities, and a user-centric design philosophy to enhance stakeholder acceptance.
The sensitivity and Monte Carlo results at the country level reinforce the aggregated findings. For Austria, P(NPV\,$>$\,0)\,$>$\,99.5\%, reflecting the resilience of a large-fleet, high-price-level context to parameter perturbations. For Hungary, P(NPV\,$>$\,0)\,$>$\,99.0\%, with the slight reduction versus Austria attributable to greater uncertainty in the solar curtailment trajectory and market price projections. For Slovenia, P(NPV\,$>$\,0)\,$>$\,98.5\%, reflecting the higher proportional weight of cross-border logistics assumptions in the benefit model. Across all three countries, BCR remains above 1.0 in more than 95\% of Monte Carlo trials, confirming broad-based investment robustness at the national scale. The country-specific diurnal profiles (Fig.~\ref{diurnal_profile}) and projections (Fig.~\ref{projections}) provide the physical basis for these national differences: Hungary's rapidly growing RES capacity and high MEF during fossil dispatch hours make the AEC and CO$_2$ streams particularly sensitive to adoption-rate assumptions, while Slovenia's cross-border logistics value is most sensitive to ET fleet growth projections. 
 
\begin{figure}[htbp]
\centering
\includegraphics[width=0.95\linewidth]{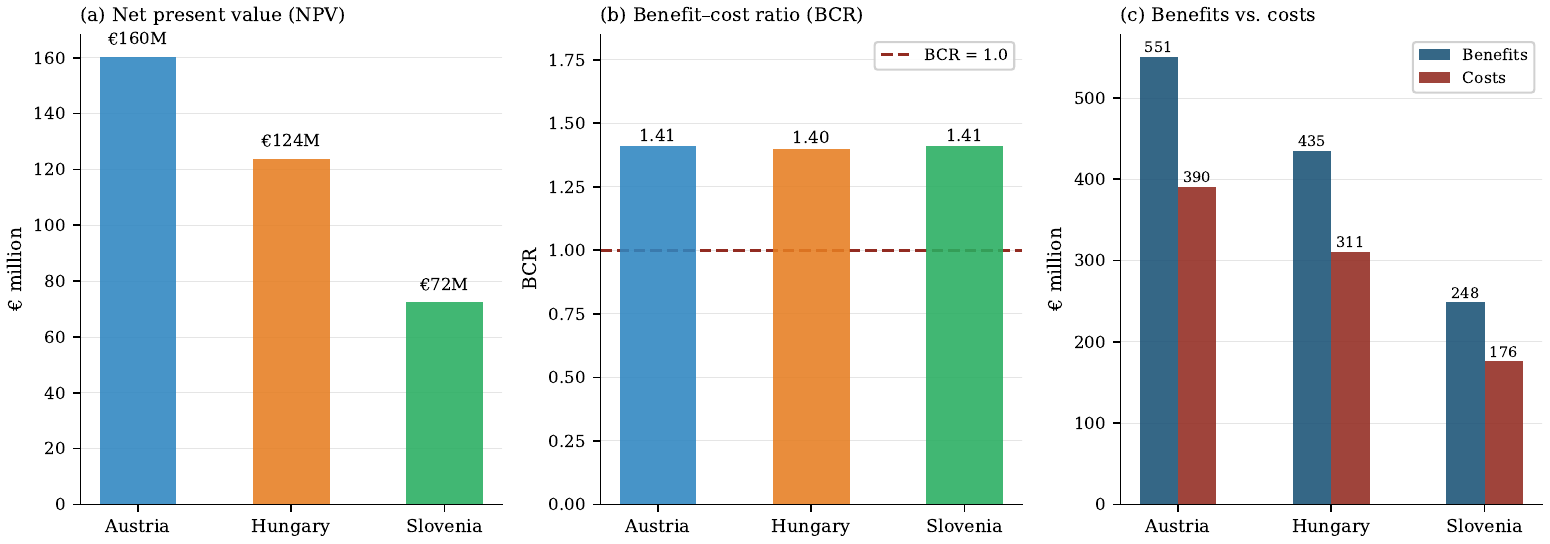}
\caption{Country-level CBA results for Austria, Hungary, and Slovenia. Austria achieves the highest absolute NPV driven by its larger fleet; all three countries maintain BCR $>$ 1.36, demonstrating consistent economic viability across diverse national contexts.}
\label{comparison}
\end{figure}

\begin{table}[htbp]
\centering
\begin{threeparttable}
\caption{Summary of CBA results and Monte Carlo simulation outcomes per country for the AI-driven ODP (PV, € million), 2026--2035.}
\label{tab_app_country_mcs_populated}
\begin{tabular}{@{}lccc@{}}
\toprule
\textbf{Metric} & \textbf{Austria (AT)} & \textbf{Hungary (HU)} & \textbf{Slovenia (SI)} \\
\midrule
\multicolumn{4}{l}{\textbf{Deterministic results (Base Case)}} \\
Total benefits (PV)         & 550.8 & 434.7 & 248.4  \\
Total costs (PV)            & 390.4 & 310.8 & 176.0  \\
NPV (PV)                    & 160.4 & 123.9 & 72.4   \\
BCR                         & 1.41  & 1.40  & 1.41   \\
\midrule
\multicolumn{4}{l}{\textbf{Monte Carlo simulation (Mean values from 10{,}000 Trials)}} \\
Mean NPV (PV)               & 168.2 & 127.5 & 74.3   \\
Mean BCR                    & 1.43  & 1.42  & 1.42   \\
$P(\mathrm{NPV} > 0)$       & $>99.5\%$ & $>99.0\%$ & $>98.5\%$ \\
$P(\mathrm{BCR} > 1)$       & $>99.5\%$ & $>99.0\%$ & $>98.5\%$ \\
\midrule
\multicolumn{4}{l}{\textbf{Dominant benefit categories (\% of national total benefits)}} \\
Economic (ROD, AEC, FES, ROETAS, CSDR-PLR) & $83.6\%$ & $81.2\%$ & $79.9\%$ \\
Environmental (CO$_2$, RAP) & $6.3\%$ & $7.0\%$ & $6.6\%$ \\
Reliability (GSMS)          & $10.1\%$ & $11.8\%$ & $13.5\%$ \\
\bottomrule
\end{tabular}
\begin{tablenotes}
\footnotesize
\item Note: These numerical values are derived from detailed country-specific financial and operational models, reflecting the specific growth rates, market prices, and grid characteristics outlined in Table \ref{tab_country_assumptions_revised} and other appendices. The slight variations in BCR across countries reflect the unique cost structures and benefit profiles of each national context. Dominant benefit categories highlight the primary drivers of value in each country, influenced by the respective characteristics of their energy and transport systems.
\end{tablenotes}
\end{threeparttable}
\end{table}

Monte Carlo simulation results consistently demonstrate a very high probability of positive NPV and BCR~$>$~1 for all three countries, confirming investment robustness across diverse national contexts. The benefit distribution varies by country. Austria shows higher absolute economic benefits, driven by its larger EV/ET stock and higher wholesale energy prices. Hungary benefits most from avoided RES curtailment, reflecting its ambitious solar expansion trajectory. Slovenia shows strong relative benefits from optimized cross-border ET logistics and effective grid-flexibility utilization.

\section{Discussion}\label{dis}
\subsection{Key findings and insights}

The three-country application yields two closely related findings. First, under the baseline assumptions, investment in the ODP is economically attractive: the aggregated NPV reaches \euro{}356.7M and the BCR is 1.41, implying that each euro invested generates \euro{}1.41 in discounted value over 2026--2035. Second, this headline result is not driven by a single dominant benefit stream. Rather, eight structurally distinct channels---spanning economic optimization, grid reliability, and environmental externalities---contribute jointly, with each linked to a specific and traceable ODP function. Figure~\ref{tornado_sensitivity} illustrates the robustness of this result: even under the most adverse single-parameter perturbation considered, namely a 10\% reduction in AI accuracy, the NPV remains positive. A central methodological insight is that coordination at the architectural level expands the valuation boundary. Cross-sectoral functions, such as forecast-informed flexibility dispatch, simultaneously reduce imbalance costs (ROETAS), flatten peak demand (CSDR-PLR), and avoid RES curtailment (AEC)---effects that siloed assessments would assign to separate systems and therefore likely undervalue. By mapping ODP functions to measurable KPI deltas prior to monetization, the framework renders these cross-stream effects traceable rather than implicit.

The quantified results further identify the integrated AI engine as the principal driver of value creation. ROETAS and CSDR-PLR together account for 25.8\% of total benefits, a contribution directly attributable to the platform's price and imbalance forecasting capabilities and its optimization of flexible-demand dispatch. The diurnal load profiles (Fig.~\ref{diurnal_profile}) clarify the underlying mechanism in physical terms: the ODP consistently shifts charging demand from high-MEF, high-price peak periods to low-MEF, low-price overnight periods, thereby capturing both market-based revenues and emissions-related co-benefits that would remain invisible in a single-sector assessment.

Country-level disaggregation (Table~\ref{tab_app_country_mcs_populated} and Fig.~\ref{comparison}) confirms investment viability across heterogeneous national contexts, although for structurally different reasons. In Austria, the results are dominated by economic benefit streams associated with its comparatively large fleet and higher price levels. In Hungary, the benefit composition is more strongly weighted toward AEC, consistent with its rapidly expanding solar capacity. In Slovenia, the comparative advantage lies in ET logistics and cross-border grid services. The \euro{}348M in GSMS benefits further underscores the ODP's role as a non-wires alternative. Environmental benefits (\euro{}128M) are also strengthened by the ODP's ability to align charging with low-carbon grid conditions, an effect that is most pronounced in Hungary because of its high-MEF fossil-dispatch hours (Fig.~\ref{diurnal_profile}).

From an implementation perspective, the sensitivity and Monte Carlo analyses indicate that delivery risk is concentrated primarily in operational execution rather than in the accounting framework itself. AI forecasting accuracy and EV/ET adoption rates consistently emerge as the most influential parameters. Accordingly, investment in data-quality infrastructure, model monitoring, and retraining pipelines should be understood not merely as a technical requirement, but as a core economic determinant of realized NPV.

\subsection{Limitations and implementation challenges}

Although the results remain robust across the tested parameter space, four limitations bound the analysis. \textit{First}, AI performance parameters---including forecast MAPE, optimization efficiency, and peak-reduction yield---are calibrated from literature benchmarks rather than from operational data generated by a deployed ODP.

\textit{Second}, the framework assumes continuous access to high-quality data across mobility, market, and grid interfaces. Intermittent data feeds or low-quality telemetry would weaken KPI improvements across multiple benefit streams simultaneously, creating a cascade risk that is not fully captured in the current sensitivity scenarios. \textit{Third}, user behavior---including EV/ET participation in demand-response programmes and willingness to cede charging flexibility---is represented through scenario-level assumptions rather than through primary behavioral estimation, such as discrete-choice experiments. \textit{Fourth}, the 10-year analytical horizon is appropriate for digital-platform upgrade cycles, but it may not fully capture the longer-lived value of grid-adjacent hardware with depreciation cycles of 20--25 years. Horizon sensitivity analysis for selected hardware-intensive components would therefore strengthen the long-run investment case.

Implementation outcomes are consequently shaped not only by techno-economic performance, but also by institutional and operational conditions. Three considerations are particularly important.

\textit{Valuation and behavioral uncertainty:} Several societal effects---including consumer empowerment, public-health co-benefits beyond monetized pollutant reductions, and network-level coordination externalities---remain unquantified, implying that the central NPV estimate is likely conservative. More importantly, the framework assumes rational uptake of incentives by EV/ET operators and fleet managers. In practice, participation is affected by information asymmetries, bounded rationality, and operational inertia. Behavioral modeling and longitudinal field evidence would therefore materially reduce the current uncertainty intervals.

\textit{Technical robustness and data risk:} Large-scale data exchange, cloud-based operation, and cross-border connectivity introduce cybersecurity and data-governance risks that constitute genuine operational costs. Model accuracy may also drift over time as mobility patterns, RES penetration, and technology costs evolve. Effective mitigation requires privacy-preserving analytics, independent performance audits, continuous retraining with drift-detection thresholds, and operational fallbacks to rule-based dispatch when AI confidence falls below predefined levels. The quantitative resilience of such controls would, however, be estimated more reliably using evidence from small-scale operational ODP deployments.

\textit{Regulatory and institutional fit:} Scalability depends critically on alignment with national grid codes, EV tariff structures, and data-sharing regulations, all of which currently vary substantially across EU member states. Fragmented or rapidly evolving policy environments increase coordination costs and may delay deployment by several years. Phased pilot projects within regulatory sandboxes---where data-sharing obligations and liability frameworks are negotiated in advance---remain the most practical mechanism for reducing implementation friction, building institutional trust, and generating the operational evidence needed to support full-scale rollout.

\section{Conclusion}\label{conc}
The transition to electrified transport and high-penetration renewable energy creates a coordination challenge that no single-sector tool can resolve: the economic, reliability, and environmental value of an AI-driven operational digital platform becomes fully visible only when mobility, energy, and grid operations are evaluated as a coupled system. This paper develops and applies a cost-benefit analysis (CBA) framework that does exactly that.

The methodological contribution is a seven-step CBA tailored to operational digital platforms. The framework links each AI module to traceable KPI deltas and monetizes the resulting benefit streams against AI-specific capital and operating costs that conventional infrastructure appraisals systematically omit. The empirical contribution is the framework’s application across three countries with distinct electrification trajectories, showing that the investment case is positive under the modeled assumptions not only in aggregate but also at the national level---and for structurally different reasons in each country. 

The broader implication is managerial and institutional. The results identify AI forecasting performance and fleet adoption as the dominant sources of value uncertainty, rather than cost overruns. This shifts investment-governance priorities toward data quality, model monitoring, and retraining pipelines---capabilities that must be planned and budgeted from project inception rather than retrofitted after deployment. For policymakers appraising digital infrastructure under Fit for 55 and REPowerEU, the framework provides a reproducible, standards-aligned methodology that makes the economic case for ODP investment transparent, auditable, and transferable to other jurisdictions.

Future work should pursue four directions: empirical pilot deployment to replace model-calibrated AI performance assumptions with operational evidence; distribution-grid impact modeling at the feeder level to address regulatory requirements in dense urban settings; integration of additional flexible assets---such as stationary storage, smart heat pumps, and industrial interruptible loads---to expand the benefit frontier beyond the streams quantified here; and extending the time horizon to 20--25 years to capture the long-run value of grid-adjacent hardware whose depreciation cycle exceeds the digital-platform upgrade cycle assumed in this study.

\section*{Acknowledgements}

This research was supported by the BEGONIA project, which has received funding from the European Commission under grant agreement No.
01133306.

\appendix
\section{Detailed cost-benefit analysis methodology and data}
\label{app_details}
This appendix formalizes the evaluation architecture by providing further details on the methodologies and data underpinning the CBA, with the objective of ensuring full transparency and reproducibility.

\subsection{Benefit monetization formulas and key parameters}\label{app_formulas}
To explicitly link the platform's technological architecture to its economic value, the following formulas translate AI-enabled functional improvements into monetized benefit streams. All monetary values are expressed in real terms (Euro 2025), and future cash flows are discounted to their present value using an economic discount rate of 4\%. 

\subsubsection{Reduction in operational downtime (ROD) cost}\label{app_formulas_rod}
Reduced operational downtime (ROD) monetizes the avoided unserved energy resulting from AI-driven predictive maintenance and enhanced situational awareness. These capabilities accelerate fault recovery and improve overall grid reliability. The annual benefit is modeled as:
\begin{equation}\label{eq_rod_appendix}
B_{\mathrm{ROD}} = C_{\mathrm{OD\_base}}\times\bigl(1 + \Delta_{\mathrm{eff}}\bigr)\times r_{\mathrm{ODP}},
\end{equation}

where $C_{\mathrm{OD\_base}}$ is the baseline operational downtime cost, $\Delta_{\mathrm{eff}}$ is the energy self-efficiency improvement fraction, and $r_{\mathrm{ODP}}$ is the ODP-enabled downtime cost savings rate. The product $C_{\mathrm{OD\_base}}(1 + \Delta_{\mathrm{eff}})$ gives the adjusted downtime cost under improved efficiency. Multiplying by $r_{\mathrm{ODP}}$ converts this into the annual monetized benefit from avoided unserved energy, consistent with smart-grid reliability benchmarks \citep{SmartGridReliabilityBenefits2019}. Parameter values are in Table~\ref{tab_params}.

\subsubsection{Revenue from optimized energy trading and ancillary services (ROETAS)}\label{app_formulas_roetas}
This benefit captures the incremental revenues and cost savings directly attributable to the ODP’s AI forecasting and optimization algorithms operating across energy arbitrage and ancillary-service markets:
\begin{equation}\label{eq_roetas_appendix}
B_{\mathrm{ROETAS}} = \sum_{t=1}^{H_{\mathrm{yr}}}\bigl(R_{\mathrm{arbitrage},t}^{\mathrm{ODP}} - R_{\mathrm{arbitrage},t}^{\mathrm{base}}\bigr)+
\sum_{s\in S_{\mathrm{anc}}}\Bigl[\sum_{t=1}^{H_{\mathrm{yr}}}\bigl(\mathrm{RFS}_{s,\mathrm{ODP},t} - \mathrm{CSP}_{s,\mathrm{ODP},t}\bigr)
  -\sum_{t=1}^{H_{\mathrm{yr}}}\bigl(\mathrm{RFS}_{s,\mathrm{base},t} - \mathrm{CSP}_{s,\mathrm{base},t}\bigr)\Bigr],
\end{equation}
where:
\begin{itemize}
\item $R_{\mathrm{arbitrage},t}^{\mathrm{ODP}}$ and $R_{\mathrm{arbitrage},t}^{\mathrm{base}}$ are the \emph{hourly} energy–arbitrage revenues (in~€) in hour $t$ with and without ODP, respectively. We compute hourly arbitrage revenue as
$$R_{\mathrm{arbitrage},t} = Q_{\mathrm{flex},t}\,\Delta P_{\mathrm{market},t}\eta_{\mathrm{cycle}},$$
with $Q_{\mathrm{flex},t}$ the flexible energy shifted in hour $t$ (MWh), $\Delta P_{\mathrm{market},t}$ the realized buy/sell price spread (€/MWh), and $\eta_{\mathrm{cycle}}\in(0,1]$ the round-trip efficiency.
\item $S_{\mathrm{anc}}=\{\mathrm{FCR},\mathrm{aFRR},\mathrm{mFRR},\mathrm{Capacity},\dots\}$ denotes the set of ancillary-service products. For each $s\in S_{\mathrm{anc}}$ and hour $t$, $\mathrm{RFS}_{s,\mathrm{ODP},t}$ and $\mathrm{CSP}_{s,\mathrm{ODP},t}$ are the ODP’s \emph{hourly} revenue and cost in~€; $\mathrm{RFS}_{s,\mathrm{base},t}$ and $\mathrm{CSP}_{s,\mathrm{base},t}$ are the corresponding baseline values.
\end{itemize}
If one prefers a compact proxy for arbitrage gains due to efficiency/forecasting improvements, the first term in \ref{eq_roetas_appendix} can be approximated by
$$f_{\mathrm{arb}}\bigl(E_{\mathrm{RES}}^{\mathrm{base}} + E_{\mathrm{EV+ET}}^{\mathrm{base}}\bigr)\,P_{\mathrm{avg}},$$
where $f_{\mathrm{arb}}$ is the net uplift fraction attributable to ODP (dimensionless), $P_{\mathrm{avg}}$ is the assumed average electricity price (€/MWh), $E_{\mathrm{RES}}^{\mathrm{base}} = C_{\mathrm{RES}}\times \mathrm{FLH}_{\mathrm{RES}}$ is baseline annual renewable output (MWh) with installed capacity $C_{\mathrm{RES}}$ (MW) and full-load hours $\mathrm{FLH}_{\mathrm{RES}}$ (h/yr), and $E_{\mathrm{EV+ET}}^{\mathrm{base}} = P_{\mathrm{EV+ET}}\times H_{\mathrm{yr}}$ is baseline annual EV+ET energy (MWh) given aggregate charging capacity $P_{\mathrm{EV+ET}}$ (MW) and $H_{\mathrm{yr}}=8760$ h/yr.

\subsubsection{Cost savings from demand response and peak load reduction (CSDR‐PLR)} \label{app_formulas_csdrplr}
System peak-demand reductions and associated avoided capacity costs are realized through the ODP’s AI-orchestrated demand response (DR) and automated load-shifting. The annual benefit is modeled as:
\begin{equation}\label{eq_csdrplr_appendix}
\begin{split}
B_{\mathrm{CSDR\text{-}PLR}} 
&= 
c_{\mathrm{DR}} \times \bigl(C_{\mathrm{RES}}\times r_{\mathrm{RES}}\times \mathrm{CF} + P_{\mathrm{EV}}\times r_{\mathrm{EV}}\times H_{\mathrm{yr}} + P_{\mathrm{ET}}\times r_{\mathrm{ET}}\times H_{\mathrm{yr}}\bigr) \\
&\quad
+\,r_{\mathrm{PL}} \times r_{\mathrm{peak}} \times \bigl(E_{\mathrm{RES}}^{\mathrm{base}} + E_{\mathrm{EV}}^{\mathrm{base}} + E_{\mathrm{ET}}^{\mathrm{base}}\bigr),
\end{split}
\end{equation}
where the first term represents benefits from cost savings resulting from demand response, and the second term represents benefits from peak-load revenue.
$c_{\mathrm{DR}}$ [€/MWh] is the unit cost saving per shifted MWh from demand response. $C_{\mathrm{RES}}$ [MW] is installed RES capacity; $\mathrm{CF}$ [h/yr] is its full-load hours, so $C_{\mathrm{RES}}\times \mathrm{CF}$ is the annual RES generation baseline (MWh). $r_{\mathrm{RES}}$ is the fraction of that output shiftable via ODP-enabled flexibility. $P_{\mathrm{EV}}$ [MW] and $P_{\mathrm{ET}}$ [MW] are the aggregated EV and ET charging capacities; $P_{\mathrm{EV}}\times r_{\mathrm{EV}}\times H_{\mathrm{yr}}$ and $P_{\mathrm{ET}}\times r_{\mathrm{ET}}\times H_{\mathrm{yr}}$ are the respective annual baseline energies (MWh) multiplied by the ODP-enabled shiftable fractions $r_{\mathrm{EV}}$ and $r_{\mathrm{ET}}$.

In the second term, $r_{\mathrm{PL}}$ [€/MWh] denotes the peak‐load reduction (PLR) revenue per shifted MWh, and $r_{\mathrm{peak}}$ is the fraction of total annual energy that can be shifted away from peak hours. We also have:
\begin{equation}
\begin{split}
E_{\mathrm{RES}}^{\mathrm{base}} &= C_{\mathrm{RES}}\times \mathrm{CF},\\
E_{\mathrm{EV}}^{\mathrm{base}}  &= P_{\mathrm{EV}}\times H_{\mathrm{yr}},\\
E_{\mathrm{ET}}^{\mathrm{base}}  &= P_{\mathrm{ET}}\times H_{\mathrm{yr}},
\end{split}
\end{equation}
where $r_{\mathrm{peak}}\bigl(E_{\mathrm{RES}}^{\mathrm{base}} + E_{\mathrm{EV}}^{\mathrm{base}} + E_{\mathrm{ET}}^{\mathrm{base}}\bigr)$ is the annual volume (MWh) shifted out of peak hours. Thus, the second term represents revenue from peak‐load reduction at peak‐period prices. 

\subsubsection{Avoided energy curtailment (AEC)}
\label{app_formulas_aec}
This formulation monetizes the value of renewable energy that is utilized rather than curtailed, a direct outcome of the AI engine's high-resolution forecasting and dynamic load-matching capabilities:

\begin{equation}\label{eq_aec_appendix}
B_{\mathrm{AEC}}=\text{RES}_{\mathrm{curt}}\times C_{\mathrm{RES}}\times\mathrm{CF}\times c_{\mathrm{curt}},
\end{equation}

where $\text{RES}_{\mathrm{curt}}$ is the fraction of RES curtailment avoided, $C_{\mathrm{RES}}$ [MW] is the installed RES capacity, and $c_{\mathrm{curt}}$ [€/MWh] is the unit cost associated with avoided curtailment. Thus, $C_{\mathrm{RES}}\times\mathrm{CF}$ gives the annual baseline RES generation (MWh), multiplying by $\text{RES}_{\mathrm{curt}}$ yields the avoided MWh, and multiplying by $c_{\mathrm{curt}}$ converts this volume to euros.

\subsubsection{Fleet energy and operational savings (FES)}
\label{app_formulas_fes}
The ODP’s AI-optimized routing algorithms (e.g., E-VRPTW solvers) and health-aware charging schedules lower operational costs by extending battery lifespans and improving net energy efficiency. The aggregated annual savings are calculated as:
\begin{equation}\label{eq_fes_appendix}
B_{\mathrm{FES}} = r_{\mathrm{fes}}\times\Bigl(N_{\mathrm{EV}}\times \eta_{\mathrm{EV}}\times S_{\mathrm{EV}} + N_{\mathrm{ET}}\times \eta_{\mathrm{ET}}\times S_{\mathrm{ET}}\Bigr),
\end{equation}
where $r_{\mathrm{fes}}$ is the overall ODP-enabled savings fraction; $N_{\mathrm{EV}}$ and $N_{\mathrm{ET}}$ are fleet sizes; $\eta_{\mathrm{EV}}$ and $\eta_{\mathrm{ET}}$ are efficiency factors capturing reduced energy consumption and extended battery life; and $S_{\mathrm{EV}}$, $S_{\mathrm{ET}}$ are annual per-vehicle savings under ODP operation. The factor $r_{\mathrm{fes}}$ consolidates two AI-driven mechanisms: battery-health optimization (8--12\% degradation reduction via SoC and C-rate management \citep{maheshwari2020optimizing}) and E-VRPTW routing efficiency (3--5\% energy savings per trip). A value of $r_{\mathrm{fes}}=0.01$ reflects a conservative 1\% net reduction in total fleet operating cost.

\subsubsection{Grid stability and management cost savings (GSMS)}
\label{app_formulas_gsms}
Grid stability and management cost savings accrue to DSOs and TSOs through improved flexibility and predictive congestion management. These benefits are quantified via reductions in expected energy not served and avoided congestion interventions. Baseline reliability is estimated from national outage statistics and grid development reports for Austria, Hungary, and Slovenia \cite{StatistaEV2023}. The platform's impact is modeled as the incremental reduction in unserved energy, driven by superior RES integration and load shifting. Monetization is anchored to country-specific Values of Lost Load (VoLL) reported in regulatory studies. The annual GSMS benefit is calculated as:
\begin{equation}\label{eq_gsms_appendix}
\begin{aligned}
B_{\mathrm{GSMS}} &=
r_{\mathrm{gsms}} \times \bigl(N_{\mathrm{EV}} \times E_{\mathrm{EV,yr}} + N_{\mathrm{ET}} \times E_{\mathrm{ET,yr}}\bigr) \times c_{\mathrm{fuel}}
+
C_{\mathrm{OD\_base}} \times C_{\mathrm{RES}} \times \mathrm{CF} \\
&\quad
+
r_{\mathrm{gsstab}} \times \bigl(N_{\mathrm{EV}} \times E_{\mathrm{EV,yr}} + N_{\mathrm{ET}} \times E_{\mathrm{ET,yr}}\bigr) \times P_{\mathrm{market}},
\end{aligned}
\end{equation}
where $r_{\mathrm{gsms}}$ is the ODP‐enabled forecasting improvement fraction (informed by smart‐grid pilot results \citep{SmartGridReliabilityBenefits2019,GridInvestmentDeferral2020}), $N_{\mathrm{EV}}$ is the total number of EVs, $N_{\mathrm{ET}}$ is the total number of ETs, $E_{\mathrm{EV,yr}}$ [MWh/yr] and $E_{\mathrm{ET,yr}}$ [MWh/year] are the annual charging energy per EV and per ET, respectively, and $c_{\mathrm{fuel}}$ [€/MWh] is the unit fuel‐saving value. The product
$r_{\mathrm{gsms}} \times \bigl(N_{\mathrm{EV}} \times E_{\mathrm{EV,yr}} + N_{\mathrm{ET}} \times E_{\mathrm{ET,yr}}\bigr) \times c_{\mathrm{fuel}}$
captures savings from reduced fuel use due to improved timing of EV/ET charging.
The term $C_{\mathrm{OD\_base}}\,C_{\mathrm{RES}}\,\mathrm{CF}$ represents avoided operational‐downtime costs: $C_{\mathrm{OD\_base}}$ is the baseline downtime cost, $C_{\mathrm{RES}}$ [MW] is installed RES capacity, and $\mathrm{CF}$ is the RES full‐load hours, so that multiplying $C_{\mathrm{OD\_base}}$ by $C_{\mathrm{RES}}\,\mathrm{CF}$ estimates the portion of downtime cost deferral attributable to RES‐driven stability improvements.
Finally, $r_{\mathrm{gsstab}}$ is the ODP‐enabled grid‐stability savings fraction. Multiplying $r_{\mathrm{gsstab}}$ by 
$\bigl(N_{\mathrm{EV}} \times E_{\mathrm{EV,yr}} + N_{\mathrm{ET}} \times E_{\mathrm{ET,yr}}\bigr) \times P_{\mathrm{market}},$ where $P_{\mathrm{market}}$ [€/MWh] is the market‐price reduction rate, yields additional benefits from reduced imbalance and congestion charges when EV/ET charging shifts under ODP’s real‐time control.

\subsubsection{\texorpdfstring{CO$_2$}{CO2} emission reduction benefits}
\label{app_formulas_co2}

Monetized greenhouse gas (GHG) emission reductions are driven by the platform's ability to maximize RES dispatch and temporally shift EV/ET charging to low-carbon hours. To capture accurate marginal effects, the model integrates spatial and temporal granularity:

\begin{equation}\label{eq_co2_appendix_refined}
B_{\mathrm{CO2}} = 
\sum_{c}\sum_{t} 
r_{\mathrm{CO2}}(t,c)\,
\bigl(C_{\mathrm{RES},t,c}\times \mathrm{CF}_{t,c} 
+ N_{\mathrm{EV},t,c}\times \eta_{\mathrm{EV}} 
+ N_{\mathrm{ET},t,c}\times \eta_{\mathrm{ET}}\bigr)\,
\mathrm{MEF}^{\mathrm{CO2}}_{t,c}\,
\mathrm{SCC}_{c},
\end{equation}

where $r_{\mathrm{CO2}}(t,c)$ is the ODP‐enabled reduction fraction varying by time and country, 
$N_{\mathrm{EV},t,c}$ is the number of EVs in operation in country $c$ at time $t$ and 
$\eta_{\mathrm{EV}}$ [MWh/vehicle] is their average energy demand after AI‐optimized charging. 
Similarly, $N_{\mathrm{ET},t,c}$ is the number of ETs, and $\eta_{\mathrm{ET}}$ [MWh/vehicle] is their annualized energy demand. 
$C_{\mathrm{RES},t,c}\times \mathrm{CF}_{t,c}$ denotes the renewable generation (capacity multiplied by capacity factor) available in country $c$ at time $t$. 
The factor $\mathrm{MEF}^{\mathrm{CO2}}_{t,c}$ [tonne CO$_2$/MWh] is the hourly marginal emission factor, 
and $\mathrm{SCC}_{c}$ [€/tonne CO$_2$] is the country‐specific social cost of carbon. 
The summation over $t$ and $c$ captures total avoided CO$_2$ emissions. Each term is weighted by the temporally varying marginal emission intensity and the regionally differentiated social cost of carbon.

\subsubsection{Reduction in air pollutant (RAP)}
\label{app_formulas_rap}
Reductions in local air pollutants (e.g., NO$_x$, SO$_x$, PM$_{2.5}$) generate monetizable public health and environmental benefits. The AI engine facilitates this by aligning transport and power demand with cleaner generation profiles. This benefit is modeled as:
\begin{equation}\label{eq_rap_appendix}
\begin{aligned}
&B_{\mathrm{RAP}}
= \sum_{c}\sum_{t}\Bigl(\alpha_{\mathrm{EV},t,c} \times D_{\mathrm{EV},t,c}\times N_{\mathrm{EV},t,c}
+ \alpha_{\mathrm{ET},t,c}\times D_{\mathrm{ET},t,c}\times  N_{\mathrm{ET},t,c}\times r_{\mathrm{ODP},t,c} \\
&+\mathrm{MEF}^p_{t,c}\times C_{\mathrm{RES},t,c}\times \mathrm{CF}_{t,c}\times r_{\mathrm{dec},t,c}\times r_{\mathrm{ODP},t,c}\Bigr)\, D^p_c,
\end{aligned}
\end{equation}
where $\alpha_{\mathrm{EV},t,c}$ [kg/km] and $\alpha_{\mathrm{ET},t,c}$ [kg/km] are pollutant emission factors for EVs and ETs in country $c$ at time $t$, reflecting local fleet composition and technology. 
$D_{\mathrm{EV},t,c}$ and $D_{\mathrm{ET},t,c}$ [km/year] are annual driving distances per EV and ET, and $N_{\mathrm{EV},t,c}$ and $N_{\mathrm{ET},t,c}$ are the respective vehicle counts. 
The products $\alpha_{\mathrm{EV}}D_{\mathrm{EV}}N_{\mathrm{EV}}$ and $\alpha_{\mathrm{ET}}D_{\mathrm{ET}}N_{\mathrm{ET}}r_{\mathrm{ODP}}$ (subscripts $t,c$ suppressed for clarity) give the pollutant mass reductions (kg) attributable to AI-enabled charging and scheduling. 

The term $\mathrm{MEF}^p_{t,c}$ [kg/MWh] denotes the hourly marginal emission factor for pollutant $p$, while $C_{\mathrm{RES},t,c}$, $\mathrm{CF}_{t,c}$, $r_{\mathrm{dec},t,c}$, and $r_{\mathrm{ODP},t,c}$ capture the RES output shifted and decarbonized through ODP optimization. 

Finally, $D^p_c$ [€/kg] represents the country-specific marginal damage cost of pollutant $p$, accounting for local air quality, health impacts, and income levels. 
Summation over $t$ and $c$ ensures that spatial and temporal heterogeneity in pollution exposure and generation mix is fully reflected in the monetized benefit, improving granularity and policy relevance relative to static national averages.

\begin{table}[H]
\centering
\caption{Parameters used in benefit valuation (Austria, Hungary, Slovenia).}
\label{tab_params}
\begin{threeparttable}
\begin{tabular}{@{}l l l S[table-format=4.1] p{6.8cm}@{}}
\toprule
Name & Symbol & Unit & {Value} & Source \\
\midrule
Baseline downtime cost & $C_{\mathrm{OD\_base}}$ & M€/yr & 24.2 & National statistics. \\
Self-efficiency improvement & $\Delta_{\mathrm{eff}}$ & \% & 15.0 & Assumption (tested in sensitivity). \\
ODP downtime savings rate & $r_{\mathrm{ODP}}$ & \% & 13.0 & Assumption. \\
Avg. electricity price & $P_{\mathrm{avg}}$ & €/MWh & 25.0 & \cite{ENTSOETransparency}. \\
Arbitrage energy share & $f_{\mathrm{arb}}$ & \% of energy & 5.0 & Assumption (proxy uplift). \\
RES capacity & $C_{\mathrm{RES}}$ & MW & {} &  \cite{ENTSOETransparency}. \\
RES full-load hours & $\mathrm{FLH}_{\mathrm{RES}}$ & h/yr & 1200 &  \cite{EurostatEnergyBalances,EEAElecIntensity}. \\
EV charging capacity & $P_{\mathrm{EV}}$ & MW & {} & dataset (Table \ref{tab_annual_ev_et_stock}); national stock data per Eurostat \cite{EurostatEnergyBalances}. \\
ET charging capacity & $P_{\mathrm{ET}}$ & MW & {} & dataset (Table \ref{tab_annual_ev_et_stock}). \\
DR unit saving & $c_{\mathrm{DR}}$ & €/MWh & {} & National regulator tariff schedules. \\
Peak-shiftable share & $r_{\mathrm{peak}}$ & \% of energy & 15.0 & Assumption. \\
Avoided curtailment share & $\mathrm{RES}_{\mathrm{curt}}$ & \% of RES & 1.0 & Assumption. \\
Curtailment valuation & $c_{\mathrm{curt}}$ & €/MWh & 50.0 & Assumption (opportunity cost proxy). \\
Fleet savings factor & $r_{\mathrm{fes}}$ & \% & 1.0 & Conservative assumption. \\
Per-EV annual saving & $S_{\mathrm{EV}}$ & €/veh$\cdot$yr & 800 & IEA Global EV Outlook \cite{IEAGEVO2024}. \\
Per-ET annual saving & $S_{\mathrm{ET}}$ & €/veh$\cdot$yr & 1920 &  \cite{ICCTZETCO}. \\
Grid-stability improvement & $r_{\mathrm{gsms}}$ & \% & 5.0 & \cite{badanjak2021distribution}. \\
Fuel value (avoided) & $c_{\mathrm{fuel}}$ & €/MWh & 500 & \cite{ACERMMR2023}. \\
Grid-stability savings rate & $r_{\mathrm{gsstab}}$ & \% & 1.0 & Assumption. \\
Market price (stability) & $P_{\mathrm{market}}$ & €/MWh & 30.0 & \cite{ACERMMR2023}. \\
Value of lost load & $\mathrm{VoLL}$ & €/MWh & {} & CEPA study for ACER \cite{CEPA2018VoLL}. \\
CO$_2$ reduction factor & $r_{\mathrm{CO2}}$ & \% & 12.0 & Assumption. \\
Hourly marginal CO$_2$ factor & $\mathrm{MEF}^{\mathrm{CO2}}_{t,c}$ & tCO$_2$/MWh & {} &  \cite{EEAElecIntensity}. \\
Social cost of carbon & $\mathrm{SCC}_{c}$ & €/tCO$_2$ & {} & \cite{EUCarbonValuation2022}. \\
Pollutant emission factor & $\mathrm{EF}_{\mathrm{poll}}$ & kg/MWh & {} &  \cite{EMEPGuidebook}. \\
Pollutant unit cost & $C_{\mathrm{poll}}$ & €/kg & {} & \cite{EEADamageCosts}. \\
\bottomrule
\end{tabular}
\end{threeparttable}
\end{table}

\subsection{CAPEX breakdown}\label{app_capex}
Table~\ref{tab_app_capex_populated} presents the discounted present value of CAPEX over 2026--2035, aggregated for Austria, Hungary, and Slovenia. Two categories are distinguished. The first covers incremental hardware investment (scaling year-over-year with fleet and RES growth) for expanding sensor networks. The second covers foundational, one-time expenditures for the core technology stack, including the primary development of AI models and dedicated computing infrastructure for their training and initial deployment.

\begin{table}[htbp]
\centering
\begin{threeparttable}
\caption{CAPEX breakdown for the AI-driven ODP (€ million, discounted total PV over 10 years, aggregated for AT, HU, SI)}
\label{tab_app_capex_populated}
\begin{tabular}{@{}p{0.83\linewidth}r@{}}
\toprule
\textbf{Cost category} & \textbf{Total (PV)} \\
\midrule
\textbf{Hardware (sensors and edge devices - incremental)} & \\
Incremental sensors for new EVs & 280.00 \\
Incremental sensors for new ETs & 120.00 \\
Incremental sensors for new RES installations & 71.65 \\
Subtotal hardware (incremental) & 471.65 \\
\midrule
\textbf{Core technology platform (initial investments and upgrades)} & \\
Adaptive environmental sensor arrays (initial deployment) & 20.00 \\
Weather forecasting sensor infrastructure & 1.20 \\
Connectivity infrastructure (5G/LPWAN) & 3.00 \\
Communication protocols & 3.00 \\
Message queue systems & 3.00 \\
Databases and data lakes (setup) & 3.00 \\
Data integration middleware (FIWARE Context Broker) & 4.50 \\
API gateway and identity management gateway & 4.50 \\
Market and operator data interfaces (EPEX SPOT APIs) & 4.50 \\
AI models for grid optimization (development and initial training) & 15.00 \\
Demand response tools (integration) & 6.00 \\
Fleet routing and dispatch engine (development and integration) & 9.00 \\
Weather forecasting algorithms (development and integration) & 3.00 \\
End-user mobile application development & 0.90 \\
Charging navigation systems integration & 3.00 \\
Comprehensive analytics tools development & 3.00 \\
Administrative dashboard development & 3.00 \\
Software deployment and system integration & 3.00 \\
Subtotal core technology platform & 89.60 \\
\midrule
\textbf{Total CAPEX (PV)} & \textbf{561.25} \\
\bottomrule
\end{tabular}
\begin{tablenotes}
    \footnotesize
\item Note: Incremental sensor CAPEX reflects the cumulative discounted cost of equipping newly adopted EVs, ETs, and RES units annually with ODP-compatible sensors and edge devices. Core technology platform costs represent foundational one-time investments for the ODP core infrastructure and software components, including dedicated AI development. These figures are based on industry benchmarks for similar digital platform deployments in the energy sector, scaled for the three-country scope. Contingencies are embedded within these estimates.
\end{tablenotes}
\end{threeparttable}
\end{table}

\subsection{OPEX breakdown}\label{app_opex}
Table~\ref{tab_app_opex_populated} presents the discounted present value of recurring annual OPEX over the 2026--2035 period, aggregated across the three countries. The annual average PV is derived from the total 10-year discounted OPEX and provides an annualized cost perspective. Costs related to sustaining and evolving the AI engine are explicitly itemized.

\begin{table}[htbp]
\centering
\begin{threeparttable}
\caption{OPEX breakdown for the AI-driven ODP (annual average PV over 10 years, € million, aggregated for AT, HU, SI)}
\label{tab_app_opex_populated}
\begin{tabular}{@{}p{0.73\linewidth}r@{}}
\toprule
\textbf{Cost category} & \textbf{annual Avg. PV} \\
\midrule
Cloud computing (infrastructure, VMs, storage, networking) & 3.9 \\
AI inference and data processing (real-time predictions, model serving) & 3.9 \\
Software maintenance and updates (platform, third-party libraries) & 2.2 \\
AI model maintenance, retraining, and performance monitoring & 2.5 \\
Grid and sensor maintenance (upkeep of physical assets) & 2.8 \\
Cybersecurity and threat monitoring (platform-wide) & 3.6 \\
Customer support and data analytics (stakeholders and users) & 1.9 \\
TSO/DSO and energy market integration (data feeds, API costs) & 2.8 \\
Grid expansion and new EV/ET integrations (operational, excl. CAPEX) & 3.2 \\
Network maintenance (5G/LPWAN) & 3.0 \\
Market trading fees (transactions, commissions) & 2.4 \\
Miscellaneous costs (training, compliance, licenses) & 2.8 \\
Labor costs (technical staff, data scientists, operational support) & 3.0 \\
\midrule
\textbf{Total annual average OPEX (PV)} & \textbf{31.6} \\
\textbf{Total 10-Year OPEX (PV)} & \textbf{315.9} \\
\bottomrule
\end{tabular}
\begin{tablenotes}
\footnotesize
\item \textit{Note:} The Annual Average PV is derived from the total 10-year discounted OPEX. The distribution across categories is proportionally guided by detailed first-year operational cost assessments, encompassing all recurring expenses for platform operation, maintenance, and support. AI-specific OPEX (e.g., Cloud computing for AI inference, AI model maintenance) accounts for approximately 20\% of total OPEX, crucial for sustaining the AI engine's performance. Contingencies are embedded within these estimates.
\end{tablenotes}
\end{threeparttable}
\end{table}

To increase transparency, AI-related expenditures are separated into one-time CAPEX (Table~\ref{tab_app_capex_populated}) and recurring OPEX (Table~\ref{tab_app_opex_populated}). CAPEX covers the upfront development and integration of AI modules; OPEX covers recurring costs including cloud hosting, AI inference and retraining, cybersecurity, and customer analytics. Both tables explicitly embed AI-specific cost items, consistent with procurement benchmarks from AWS, Azure, Eurostat ICT reports, and IEA/ACER market integration studies.
Table~\ref{tab_ai_costs} summarizes unit cost assumptions for AI model training, cloud infrastructure, and maintenance. These values draw on European and international benchmarks, scaled using the adoption and growth assumptions of Appendix~A2. Together, AI-specific expenditures account for approximately 20\% of total OPEX and 15\% of total CAPEX.

\begin{table}[H]
\centering
\caption{AI-specific cost assumptions (baseline 2026 values, aggregated across AT, HU, SI).}
\label{tab_ai_costs}
\begin{tabular}{@{}p{0.45\linewidth}p{0.25\linewidth}p{0.25\linewidth}@{}}
\toprule
\textbf{AI function / resource} & \textbf{Unit cost assumption} & \textbf{Source} \\
\midrule
AI model training (grid + transport) & €1-1.2 million per comprehensive model & \citep{AWSpricing2024,Azurepricing2024} \\
AI inference (real-time predictions) & €0.05–0.07 per CPU-hour & \citep{AWSpricing2024,Azurepricing2024} \\
Cloud storage and integration & €0.02–0.03 per GB-month & \citep{EurostatICT2023} \\
Model retraining and monitoring & €1.2–1.5M per model-year & \citep{ECAI2023,IEA2023grids} \\
Cybersecurity and threat monitoring & €200–300k per country-year & \citep{IEA2023grids} \\
API/market integration & €1.0–1.3M per country-year & \citep{ACER2023market} \\
Labor (AI/data engineers, ops staff) & €45–60k per FTE per year & \citep{EurostatWages2023} \\
\bottomrule
\end{tabular}
\end{table}

\section{Annual projections and underlying data for the ODP case studies}
\label{app_annual_proj}

This appendix provides annual projections for key parameters, monetized benefits, and costs across Austria, Hungary, and Slovenia. These projections form the empirical basis for the discounted present values reported in the main paper, derived from the modeling framework and assumptions of Section~\ref{meth}.

\subsection{Annual EV/ET Stock and RES capacity projections}
EV/ET adoption rates and RES capacity growth are fundamental macroeconomic drivers of the ODP's scale of benefits and costs. Tables~\ref{tab_annual_ev_et_stock} and \ref{tab_annual_res_capacity} present annual projected fleet stocks and installed RES capacity for each country over 2026--2035. Projections are consistent with national energy and climate plans (NECPs), established industry forecasts, and historical trends as referenced in Table~\ref{tab_country_assumptions_revised}.

\begin{table}[H]
\centering
\footnotesize
\caption{Annual projected EV and ET stock (thousands) for AT, HU, and SI (2026--2035)}
\label{tab_annual_ev_et_stock}
\begin{tabular}{@{}c r r r r r r@{}}
\toprule
\textbf{Year} & \multicolumn{3}{c}{\textbf{EV Stock (thousands)}} & \multicolumn{3}{c}{\textbf{ET stock (thousands)}} \\
\cmidrule(lr){2-4} \cmidrule(lr){5-7}
& Austria & Hungary & Slovenia & Austria & Hungary & Slovenia \\
\midrule
2026 & 343.8 & 100.8 & 50.0 & 4.4 & 0.4 & 0.3 \\
2027 & 398.8 & 118.8 & 57.7 & 4.9 & 0.5 & 0.3 \\
2028 & 462.6 & 140.0 & 66.6 & 5.5 & 0.5 & 0.4 \\
2029 & 536.6 & 165.1 & 76.9 & 6.2 & 0.6 & 0.5 \\
2030 & 622.5 & 194.7 & 88.8 & 7.0 & 0.7 & 0.5 \\
2031 & 722.1 & 229.6 & 102.6 & 7.9 & 0.8 & 0.6 \\
2032 & 837.7 & 270.6 & 118.5 & 8.9 & 0.9 & 0.7 \\
2033 & 971.7 & 318.9 & 136.9 & 10.1 & 1.0 & 0.8 \\
2034 & 1127.2 & 375.9 & 158.2 & 11.4 & 1.1 & 0.9 \\
2035 & 1307.5 & 443.3 & 182.7 & 12.8 & 1.3 & 1.0 \\
\bottomrule
\end{tabular}
\end{table}

\begin{table}[H]
\centering
\footnotesize
\caption{Annual projected RES capacity (Solar + Wind, GW) for AT, HU, and SI (2026--2035)}
\label{tab_annual_res_capacity}
\begin{tabular}{@{}c r r r@{}}
\toprule
\textbf{Year} & \textbf{Austria (GW)} & \textbf{Hungary (GW)} & \textbf{Slovenia (GW)} \\
\midrule
2026 & 10.5 & 8.0 & 1.5 \\
2027 & 11.2 & 10.7 & 1.6 \\
2028 & 11.8 & 13.5 & 1.8 \\
2029 & 12.5 & 16.3 & 1.9 \\
2030 & 13.2 & 19.1 & 2.1 \\
2031 & 13.9 & 21.9 & 2.2 \\
2032 & 14.6 & 24.7 & 2.4 \\
2033 & 15.3 & 27.5 & 2.5 \\
2034 & 16.0 & 30.3 & 2.7 \\
2035 & 16.7 & 33.1 & 2.8 \\
\bottomrule
\end{tabular}
\end{table}

\subsection{Annual monetized benefits of the ODP (aggregated)}
Table~\ref{tab_annual_benefits_agg} presents annual monetized benefits aggregated across Austria, Hungary, and Slovenia prior to discounting. The figures illustrate the year-on-year compounding growth of benefits as the platform achieves deeper network effects alongside rising EV/ET market penetration and expanding RES capacities.

\begin{table}[H]
\centering
\caption{Annual monetized benefits of ODP (aggregated for AT, HU, SI, € million)}
\label{tab_annual_benefits_agg}
\resizebox{\columnwidth}{!}{%
\begin{tabular}{@{}c S[table-format=3.1]S[table-format=3.1]S[table-format=3.1]S[table-format=3.1]S[table-format=3.1]S[table-format=3.1]S[table-format=3.1]S[table-format=3.1]S[table-format=3.1]@{}}
\toprule
\textbf{Year} & \textbf{ROD} & \textbf{ROETAS} & \textbf{CSDR-PLR} & \textbf{FES} & \textbf{AEC} & \textbf{GSMS} & \textbf{CO$_2$} & \textbf{RAP} & \textbf{Total annual benefits} \\
\midrule
2026 & 3.4 & 11.6 & 9.6 & 4.2 & 20.0 & 23.3 & 6.9 & 2.1 & 81.1 \\
2027 & 4.0 & 12.2 & 10.8 & 4.7 & 24.2 & 25.2 & 7.3 & 2.5 & 90.9 \\
2028 & 4.6 & 12.8 & 12.1 & 5.3 & 25.5 & 27.3 & 7.7 & 2.7 & 98.0 \\
2029 & 5.3 & 13.5 & 13.6 & 5.9 & 26.8 & 29.6 & 8.1 & 3.1 & 105.9 \\
2030 & 6.1 & 14.3 & 15.2 & 6.7 & 28.3 & 32.2 & 6.4 & 3.5 & 112.7 \\
2031 & 6.9 & 15.0 & 17.1 & 7.5 & 29.7 & 35.1 & 8.5 & 3.8 & 123.6 \\
2032 & 8.0 & 15.8 & 19.2 & 8.5 & 31.3 & 38.2 & 9.0 & 4.3 & 134.3 \\
2033 & 9.3 & 16.7 & 21.6 & 9.6 & 32.9 & 41.6 & 9.5 & 4.8 & 146.0 \\
2034 & 10.6 & 17.5 & 24.3 & 10.8 & 34.6 & 45.5 & 10.0 & 5.5 & 148.8 \\
2035 & 12.2 & 18.5 & 27.3 & 12.2 & 36.5 & 49.7 & 11.1 & 6.1 & 173.6 \\
\bottomrule
Total & 70.6 & 148.2 & 170.8 & 75.6 & 292.9 & 347.9 & 89.0 & 38.6 & 1233.9 \\
\bottomrule
\end{tabular}%
}
\end{table}

\subsection{Annual costs of the ODP (aggregated)}
Table~\ref{tab_annual_costs_agg} details annual CAPEX and OPEX aggregated across Austria, Hungary, and Slovenia prior to discounting. The figures illustrate the lifecycle investment profile: intensive upfront capital allocation for platform architecture in the first year, followed by stable and declining recurring operating expenditures.

\begin{table}[H]
\centering
\footnotesize
\begin{threeparttable}
\caption{Annual costs of ODP (aggregated for AT, HU, SI, undiscounted € million)}
\label{tab_annual_costs_agg}
  \begin{tabular}{@{}c S[table-format=3.1] S[table-format=2.1] S[table-format=3.1]@{}}
    \toprule
    \textbf{Year} & \textbf{Annual CAPEX (€ million)} & \textbf{Annual OPEX (€ million)} & \textbf{Total annual costs (€ million)} \\
    \midrule
    2026 & 248.5 & 37.4 & 285.9 \\
    2027 &  17.3 & 36.0 &  53.3 \\
    2028 &  18.7 & 34.6 &  53.4 \\
    2029 &  20.3 & 33.3 &  53.6 \\
    2030 &  22.1 & 32.0 &  54.1 \\
    2031 &  24.1 & 30.8 &  54.9 \\
    2032 &  26.3 & 29.6 &  55.9 \\
    2033 &  28.7 & 28.5 &  57.1 \\
    2034 &  31.4 & 27.4 &  58.7 \\
    2035 &  34.3 & 26.3 &  60.6 \\
    \midrule
    \textbf{Total} & 471.7 & 315.9 & 787.5 \\
    \bottomrule
  \end{tabular}
\begin{tablenotes}
  \footnotesize
\item Note: Annual CAPEX and OPEX are undiscounted annual sums over the 2026–2035 period, aggregated across Austria, Hungary, and Slovenia. The \textit{Total} row shows the ten‐year aggregate costs for each category. In addition, a one‐time upfront fixed cost of €89.6 million is incurred in 2026 to cover connectivity (5G/LPWAN), communication protocols and message queues, databases and data lakes, data integration middleware, market integration middleware, AI models for grid optimization, demand response tools, route optimization engines, weather-forecasting algorithms, software deployment and integration.
\end{tablenotes}
\end{threeparttable}
\end{table}

\newpage
\bibliographystyle{elsarticle-num-names}
\newpage
\bibliography{reference.bib} 
\end{document}